\documentclass[12pt,preprint]{aastex}
\usepackage{lscape}
\usepackage{natbib}

\newcommand{\caii}{Ca~{\small II}}

\shorttitle{LMC Bar}
\shortauthors{Cole et al.}

\begin{document}

\title{Spectroscopy of Red Giants in the LMC Bar:\\
Abundances, Kinematics, and the Age-Metallicity Relation\\
}

\author{Andrew A. Cole\altaffilmark{1,2}, Eline Tolstoy}
\affil{Kapteyn Astronomical Institute, University of Groningen,
Postbus 800, 9700~AV Groningen, Netherlands;
{\it cole@astro.rug.nl, etolstoy@astro.rug.nl}}

\author{John S. Gallagher, III}
\affil{Department of Astronomy, University of Wisconsin--Madison,
5534 Sterling Hall, 475 North Charter Street, Madison, WI 53706--1582;
{\it jsg@astro.wisc.edu}}

\and

\author{Tammy A. Smecker-Hane}
\affil{Department of Physics and Astronomy, University of California
at Irvine, 4129 Frederick Reines Hall, Irvine, CA 92697--4575;
{\it tsmecker@uci.edu}}

\altaffiltext{1}{NOVA Fellow.}
\altaffiltext{2}{Visiting Astronomer, Paranal Observatory.}

\begin{abstract}
We report metallicities and radial velocities derived from
spectra at the near-infrared calcium triplet for 373 red giants
in a 200 square arcminute area at the optical center of the
LMC bar.  These are the first spectroscopic abundance measurements
of intermediate-age and old field stars in the high surface
brightness heart of the LMC.  The metallicity distribution
is sharply peaked at the median value
[Fe/H] = $-$0.40, with a small tail of stars extending down
to [Fe/H] $\leq$ $-$2.1; 10\% of the red giants are observed
to have [Fe/H] $\leq$ $-$0.7.  The relative lack of metal-poor
stars indicates that the LMC has a ``G dwarf'' problem, similar 
to the Milky Way.  The abundance distribution can be closely
approximated by two Gaussians containing 89\% and 11\% of the
stars, respectively: the first component is centered at
[Fe/H] = $-$0.37 with $\sigma$ = 0.15, and the second
at [Fe/H] = $-$1.08 with $\sigma$ = 0.46.  The dominant
population has a similar metallicity distribution to the
LMC's intermediate-age star clusters.  The mean heliocentric
radial velocity of the sample is 257 km~sec$^{-1}$, corresponding
to the same center of mass velocity as the disk (measured at 
larger radii).  Because of the central location of our field,
kinematic constraints are not strong, but there is no evidence
that the bar deviates from the general motion of the LMC disk.
The velocity dispersion of the
whole sample is $\sigma _v$ = 24.7 $\pm$0.4 km~sec$^{-1}$.  When cut by
metallicity, the most metal-poor 5\% of stars ([Fe/H] $< -$1.15)
show $\sigma _v$ = 40.8 $\pm$1.7 km~sec$^{-1}$, more than twice the value
for the most metal-rich 5\%; this suggests that an old, thicker
disk, or halo population is present.  The age-metallicity
relation (AMR) is almost flat during the period from 5--10~Gyr ago,
with an apparent scatter of $\pm$0.15 dex about the mean
metallicity for a given age.  Comparing to chemical evolution 
models from the literature, we find that a burst of star formation
3~Gyr ago does not reproduce the observed AMR
more closely than a steadily declining star-formation rate.
The AMR suggests that the epoch of enhanced star formation, if
any, must have commenced earlier, $\approx$6~Gyr ago-- the exact
time is model-dependent.  We compare
the properties of the LMC and the Galaxy, and discuss our
results in the context of models that attempt to use tidal
interactions with the Milky Way and Small Magellanic Cloud
to explain the star and cluster formation histories of the LMC.
\end{abstract}

\section{Introduction}

The Large Magellanic Cloud (LMC) is the nearest actively
star-forming galaxy to us, and an invaluable laboratory
for the study of stellar and galactic evolution.  With
a total mass of $\sim$10$^{10}$ M$_{\sun}$ \citep{wes97},
the LMC lies just at the borderline between dwarf and
giant galaxies, in a regime where the
scaling relations of basic galaxy properties (metallicity,
mean stellar age, internal structure) with mass undergo a
fundamental qualitative
change \citep{kau03}.  Additionally, the LMC is deeply affected
by the tidal forces stemming from its interactions with
the Small Magellanic Cloud and the Milky Way.  Such
interactions have been plausibly connected to major events
in the lifetime of galaxies, including the creation of bulges,
bars, and/or thick disks; and to starburst activity
\citep[e.g.][]{gar99}.

To accurately measure the histories of star formation and
chemical evolution of the LMC is a major challenge for
astrophysicists.  While general characteristics of its
evolution-- such as the relatively greater number of
stars aged a few gigayears compared to the Milky Way
\citep{but77}-- have been known for decades, the details
are only now able to be measured.  For instance, it was
not until the construction
of the cleanest possible deep color-magnitude
diagrams (CMDs) from the WFPC2 camera aboard the Hubble
Space Telescope that it became apparent that the variations
in field star formation rate have been largely decoupled
from variations in the cluster formation rate \citep{hol99}.
Even these modern analyses are not without their difficulties,
chiefly the extreme disparity in angular size between the
LMC ($>$10$^6$ square arcminutes) and the WFPC2 field of
view ($\approx$ 5 square arcminutes).  

A major factor limiting the precision of star-formation
history (SFH) measurements
based on deep CMDs is the age-metallicity degeneracy-- the
fact that age and metallicity can be played off against one
another to recreate closely similar distributions of stars
in CMDs \citep[e.g.,][]{wor99}\footnote{In principle, the 
combination of photometry of the main sequence and the red giant
branch breaks this degeneracy; in practice
distance and reddening uncertainties, the arbitrary distributions
of metallicity and age in a galaxy, and the difficulties
with theoretical models for RGB evolution make this problematic.}.
This is exacerbated by the current lack of knowledge
of the LMC's age-metallicity relation (AMR).  Star clusters, the
most obvious tracers of such a relation, are famously scarce
in the LMC for ages between 3--10~Gyr \citep{dac91,gei97}.  This age
gap spans over half the age of the Universe; it includes
the likely epoch of galactic disk formation around redshift $z$
$\approx$1--1.5, and probably spans four LMC-Milky Way 
orbital periods \citep[e.g.,][]{gar94,bek04}.  \citet{bek04}
draw a connection between the tidal capture of the Small 
Magellanic Cloud (SMC) by the LMC and the end of the cluster
age gap-- this capture event had previously been thought
unlikely \citep*[e.g.,][and references therein]{gar94}.

For these reasons we have begun to measure the chemical abundances
of the field stars in the LMC: to fill in the cluster age gap,
to measure the variation of metallicity with radius across the
LMC, and to reliably distinguish the bar, disk, and possible thick 
disk or halo populations from each other.
Because the shape
of the metallicity distribution function (MDF) is not known
{\it a priori}, it is important to measure the largest possible
sample.  The brightest common stars in
the age range from $\approx$1--14~Gyr are red giants; in the LMC,
they have magnitudes I $\geq$14.8.  Because high-dispersion spectroscopy
of faint red giants is extremely expensive in telescope time, we 
rely on spectra of moderate resolution to derive 
abundances good to $\pm$0.1--0.2 dex by comparison to star clusters
of known metallicity.  The near-infrared calcium~II triplet 
($\lambda \approx$8500 \AA ) is the most widely used such
technique.  It has been very successfully applied to LMC star
clusters, beginning with a landmark paper by \citet{ols91}
(hereafter OSSH).
The results from OSSH have become the standard reference for
abundances of LMC clusters, and have been used as the basis for
simulations of the LMC's chemical evolution \citep[e.g.,][]{pag98}
(hereafter PT98).  The cluster-based AMR has in turn been taken as 
a guide to the AMR in derivations of the star formation history
based on deep color-magnitude diagrams of the field populations
\citep[e.g.,][]{gal96,geh98,hol99,sme02}.  It was not feasible
to obtain large samples of
field star abundances prior to the advent
of efficient multi-object spectrographs, first on 4-meter 
class telescopes, and more recently at 8-meter class telescopes.

We began to obtain abundances for red giants in two fields
of the inner LMC disk using long-slit spectroscopy in 
\citet[][Paper I]{col00}, and expanded the sample six-fold
using the Hydra multiobject spectrograph at the Cerro
Tololo Inter-American Observatory 4-meter telescope
\citep[][Paper II]{sme04}.  That study found
the mean metallicity of red giants at a radius of
roughly one disk scale length to be [Fe/H] = $-$0.45,
with fewer than 10\% of the RGB stars more metal-poor
than [Fe/H] = $-$1.  We found evidence for a composite
kinematic structure of the LMC disk, indicating either
a velocity dispersion increasing with age, or a segregation
by metallicity into thin and thick disks.  Paper~II found
the inner disk MDF to differ strongly from the only 
previous field star study, which had targeted a radially
distant location \citep[projected radius $\approx$8$\arcdeg$][]{ols93},
in having a much smaller fraction of metal-poor stars and
a sharper peak around the median.  We used our abundance data
and photometry to constrain the age-metallicity relation of the
inner disk, finding a slow increase in mean abundance with time.

In this paper, we present our measurements of the
chemical abundances and radial velocities of a large
number of red giant stars in the highest surface brightness
region of the LMC: its bar.  
Bar fields could not be targeted
using Hydra because the wide (2 arcsecond) diameter of its
fibers and the difficulties of sky subtraction using fiber systems.
We begin by giving some background
on the LMC and on the field studied here; this bar field includes
the fields singled out for detailed study with WFPC2 by 
the WFPC2 team \citep{geh98,hol99} and by our guest observer program
\citep[GO7382;][]{sme99,col02,sme02}.  Section 2 describes our
observing program at the 8.2 meter Yepun (VLT-UT4)
telescope at the European Southern Observatory's Paranal
Observatory, and the derivation of metallicities and
radial velocites from our spectra.  We also discuss
how we combine optical photometry and the new metallicity
information to estimate the stellar ages of the target
red giants.  In section 3 we present the bar field
MDF, comparing to the Solar neighborhood, other regions
of the LMC, and the LMC star clusters.  We explore the
connection between kinematics and metallicity and how the
line of sight velocity dispersion changes with abundance.
Section~3.3 compares the derived AMR to the predictions
of chemical evolution models based on both smooth and
bursting histories of star formation.  Our results are
summarized and the implications discussed in Section~4.

\subsection{Maps \& Terminology}
\label{sec-map}

\begin{figure}
\plotone{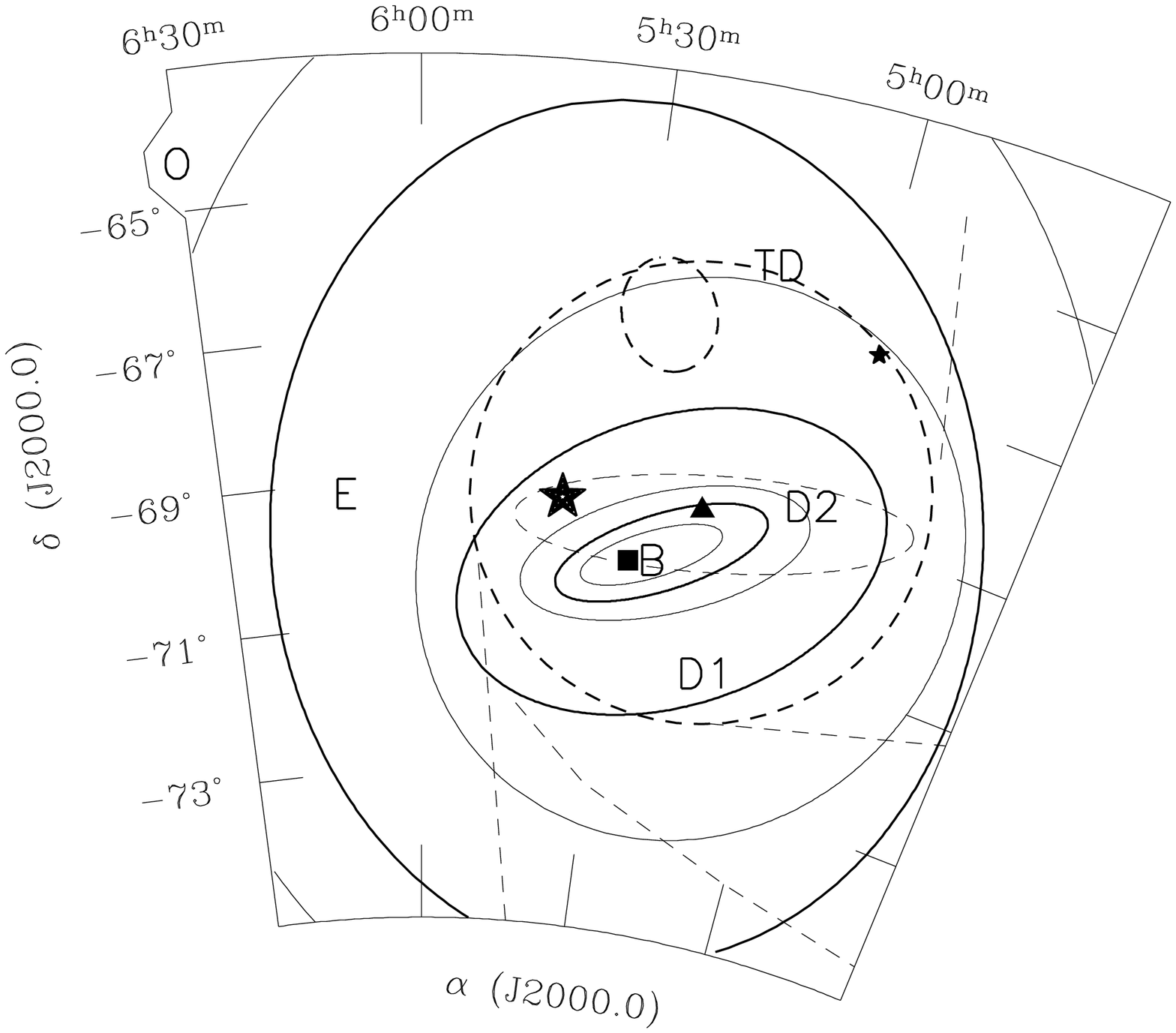}
\caption{Schematic map of the LMC, showing
near-infrared isopleths based on \citet{van01}, at semimajor
axis values $a$ = 1, 1.5, 2, 3, 4, 6, 8 degrees (solid lines).
Major large-scale \ion{H}{1} features are also sketched in,
following the maps in \citet[][dashed lines]{sta03}. The rotation
centers of intermediate-age stars \citep[][$\blacksquare$]{van02}
and \ion{H}{1} \citep[][$\blacktriangle$]{kim98} are plotted;
the LMC's two biggest \ion{H}{2} regions are shown for reference
($\bigstar$, 30~Doradus; $\star$, N11).  The alphanumeric tags 
mark areas singled out by our group for study of the field red
giant metallicity distribution; see text for details.
\label{fig-map}}
\end{figure}

To orient the reader and allow us to place our results in
a broader context, we show a schematic diagram of the LMC
in Figure~\ref{fig-map}.  The map is an Albers equal-area conic
projection\footnote{The origin is at $\alpha_0$ = 6$^{\mathrm h}$, 
$\delta_0$ =
$-$90$^{\circ}$, with reference latitudes $\delta_1$ = $-$90$^{\circ}$,
$\delta_2$ = 0$^{\circ}$.
} \citep{wei99}
of equatorial coordinates spanning $\approx$11$^{\circ}$
$\times$ 12$^{\circ}$.  The major large-scale features of the LMC
stellar and gas distributions are shown.  The solid ellipses follow the
smoothed near-infrared isopleths as fit by \citet{van01}, with semimajor
axes of 1, 1.5, 2, 3, 4, 6, and 8 degrees.  Within the 2$^{\circ}$
isopleth, the red starlight is dominated by the bar; outside this
radius, the isopleths are elongated towards the Milky Way \citep{van01},
and virtually all the stars have disk-like kinematics \citep{sch92}.
The rotational center of this disk
\citep{van02} is marked by the black square.

As many authors
have remarked, the neutral hydrogen distribution is significantly offset
from the center of the starlight; we show
the kinematic centroid of the \ion{H}{1} \citep{kim98} as a
black triangle.  Major \ion{H}{1} features are plotted with
dashed lines, following the maps in \citet{sta03}: the main
\ion{H}{1} disk, roughly 9~kpc across, the archetypal supergiant
shell LMC4 near the northern edge of the disk, and several diffuse,
tidal arms that spread out from
the southeast and the west side of the disk.  As noted by
\citet{sta03}, the \ion{H}{1} distinctly resembles a barred,
late-type spiral for velocities between 260--280 km s$^{-1}$,
with two arms connected by a bridge of gas.  We have plotted
this bridge in Figure~\ref{fig-map}, which serves to show that
there is no strict morphological relation between the optically-identified
bar and any major structure in the gas distribution.

Additional reference points are given by star symbols
marking the locations of the two most active star-forming
regions in the LMC: the 30~Doradus complex northeast of the
bar, and the N11 region near the northwest edge of the \ion{H}{1}
disk.  The distribution of stars--- particularly the most recent
generations--- and gas is incredibly complex and structured on all
scales, but this sketch is sufficient to identify the major
morphological features that bear on our results.  The Galactic center
is toward the south; in this representation, the SMC is to the lower
right, the orbit of the Clouds carries them to the northeast
(towards the Galactic plane), and the LMC's rotation is clockwise.

The features detailed above place our metallicity and kinematics
results into larger context.  We have been concerned with five
fields, marked by the appropriate alphanumeric tags in Figure~\ref{fig-map}.
Each field contains between 36 and 373 field red giants for which
spectra at the \ion{Ca}{2} triplet have been obtained.  The inner
disk fields have been studied by \citet*[][``D1'', paper~I]{col00}
and \citet[][``D1'' and ``D2'', paper~II]{sme04} using the CTIO 4-meter
telescope. Results for the transitional
disk field (``TD''), so called because of its location near the edge of the
\ion{H}{1} disk, and the eastern field (``E'') will be reported in a
future paper (Cole et al.\ 2004, in preparation).   The bar (``B'')
field is effectively located at the heart of the LMC, 0$\fdg$3 from
the center-of-rotation of carbon stars \citep{van02}.
A small number of
stars in the outer (``O'') field were observed by \citet{ols93}; these
have remained for more than a decade the only abundance measurements
of field giants lying outside the gas-rich disk.

\begin{figure}
\plotone{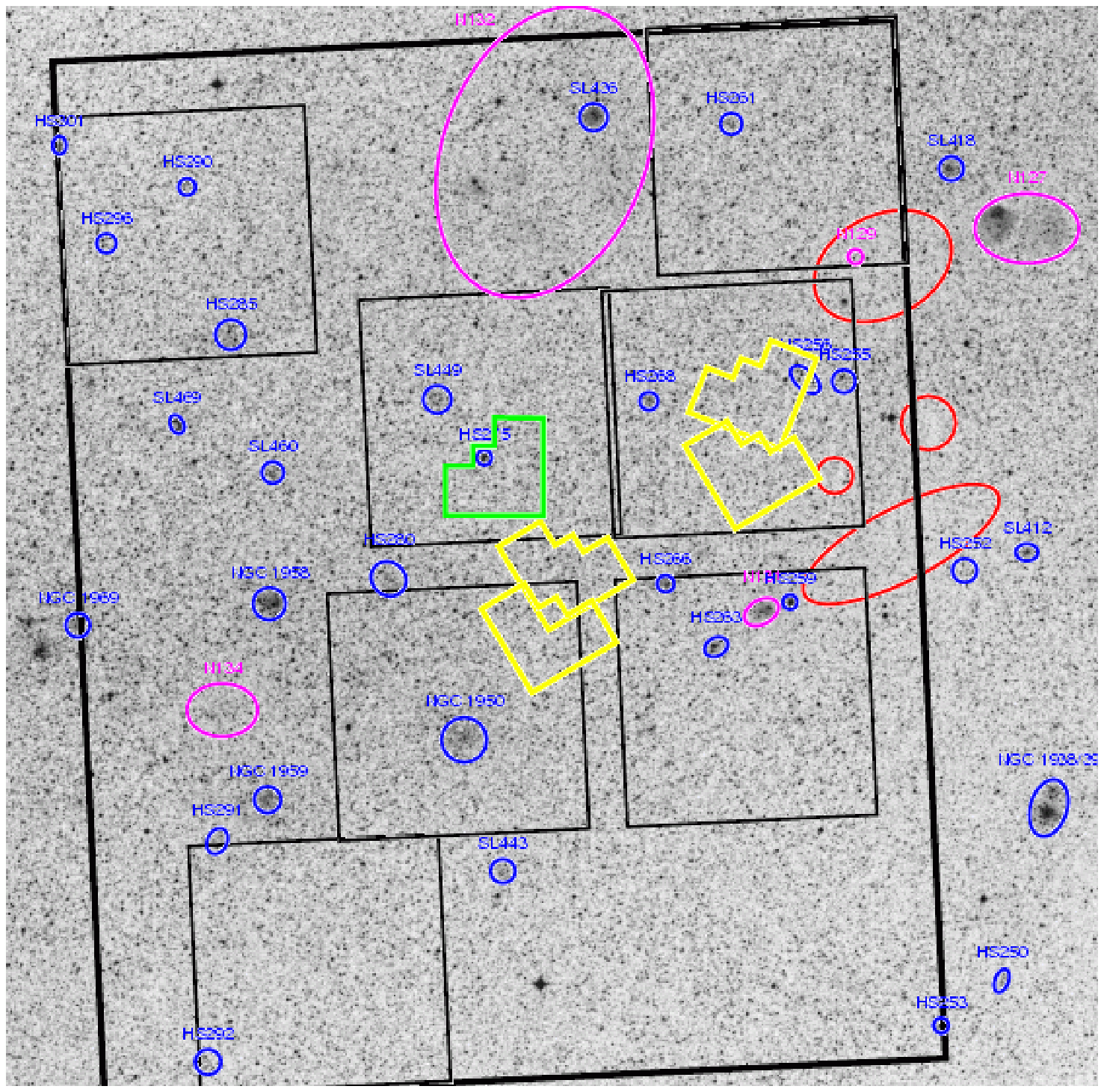}
\caption{Digitized Sky Survey image of the
bar field, centered at ($\alpha _{2000}$, $\delta _{2000}$) =
(5$^{\mathrm h}$ 24$^{\mathrm m}$, $-$69$^{\circ}$ 49$\arcmin$)
and spanning 30 arcminutes.  North is up, East is to the left.
The heavy black rectangle shows the region enclosed by
$-$70$^{\circ}$05$\arcmin$ $\leq$ $\delta$ $\leq$
$-$69$^{\circ}$35$\arcmin$, 5$^{\mathrm h}$22$^{\mathrm m}$ 
$\leq$ $\alpha$ $\leq$ 5$^{\mathrm h}$26$^{\mathrm m}$30$^{\mathrm s}$,
within which our seven FORS2 pointings (black squares) are contained.
The location of five deep WFPC2 imaging fields useful for
measuring the field star-formation history of the bar are oveplotted
in yellow \citep{sme02} and green \citep{hol99}.  Star clusters
and discrete ISM structures have been plotted and labelled
following the atlas of \citet{hod67}.  Labelled ellipses mark star
clusters (blue) and \ion{H}{2} regions (magenta).  The four red
ellipses along the eastern edge of our FORS2 fields mark the 
positions of dust clouds identified in \citet{hod72}.
\label{fig-bar}}
\end{figure}

The remainder
of this paper is concerned with the bar field.
Figure~\ref{fig-bar} shows the bar field in more detail, oriented
with North at the top and East to the left.  The
image is in the infrared band of the second Digitized Sky
Survey, obtained from the Canadian Astrophysics Data Centre.
It is centered at ($\alpha _{2000}$, $\delta _{2000}$) =
(5$^{\mathrm h}$24$^{\mathrm m}$, $-$69$^{\circ}$49\arcmin)
and spans 30 arcminutes.
Within this area are identified 30 star clusters (blue ellipses)
and five areas of H$\alpha$ emission (magenta ellipses), labelled
following the atlas of \citet{hod67}.  Four unlabelled red ellipses mark
obscuring dust clouds identified in \citet{hod72}.  OB associations
are notably absent from the field despite its high stellar density.
Minor \ion{H}{2} regions are present, but easily avoided.  The
large emission region N132 is a fossil \ion{H}{2} region around
the oxygen-rich supernova remnant N132D, which probably arose from
a Type~Ib supernova some 2500~yr ago \citep{bla00}.  N132D is injecting
dense knots of oxygen-rich material into its immediate vicinity
\citep{las78}, a vivid reminder that the abundances of the red giants
measured here are only indirectly related to the present-day gas
phase abundances.  The region around N132 contains a
few blue supergiants in projection \citep{san70};
this type of stellar population is far more common further west along
the bar and northeast towards 30~Doradus.  In many places the contribution
of intermediate-age and old stars is impossible to isolate cleanly.

Features of the diffuse interstellar medium
are omitted for readability, but the
field is comparatively simple in structure compared to much of the
LMC.  The neutral hydrogen in this field is not broken into high-
and low-velocity components, as it is in large regions to the northwest
and southeast; the \ion{H}{1} column density through this field
is roughly N$_{\mathrm{HI}}$ = 8$\times$10$^{20}$ cm$^{-2}$ \citep{luk92}.
Two relatively minor molecular clouds, identified based on
carbon monoxide line emission at 2.6~mm by \citet{coh88}, are found in the
southwest corner of the field and much of the east-central portion.

Areas included for study in this paper or related work are
also marked.  The yellow WFPC2 footprints
show the fields in which we have
obtained deep images in V and I band in order to create color-magnitude
diagrams reaching magnitude $\sim$26 \citep[][Cole et al.\ 2004,
in preparation]{sme02}. 
Similar photometric data were obtained by the WFPC2
team \citep{geh98,hol99} in the area marked by the green WFPC2 footprint.
The spectroscopic data presented in this
paper were obtained within the seven fields shown by the black squares,
each of which is 6.8 arcminutes on a side.
The heavy black rectangle outlines the region between
5$^{\mathrm h}$22$^{\mathrm m}$ $\leq$ $\alpha _{2000}$ $\leq$
5$^{\mathrm h}$26$\fm$5,
$-$70$^{\circ}$05\arcmin\ $\leq$ $\delta _{2000}$ $\leq$
$-$69$^{\circ}$35\arcmin;
this is a convenient simple border for the irregular region comprised
of tiled FORS2 fields.

Overall, the bar field is dense with stars \citep[surface brightness
$\Sigma _{\mathrm V}$ = 20.7 mag/arcsec$^2$:][]{dev72}, but is not
a region of strong current star formation or high dust obscuration.
\cite{sta03} estimate
the foreground reddening towards the bar field to be E(B$-$V) = 0.06 mag.
With the column densities of \ion{H}{1} reported in \citet{luk92}, and
applying the N$_{\mathrm{HI}}$--E(B$-$V) relation from \citet{kor82},
we make a first estimate of the mean reddening to the giants in our
field as E(B$-$V) = 0.08 $\pm$0.02 mag.  Some differential reddening
is almost certainly present, as will be discussed below.  Relatively
free of recent activity, the bar field is expected to be an ideal
place to study the intermediate-age and old stars of the central
regions of the LMC.

While the bar field is as close as practical to the centroid of old
(red) stars in the LMC, it is offset from the centroid of bright blue
stars, which are more well-aligned with the \ion{H}{1} than with the red
starlight \citep[e.g.][]{dev72}.  The flocculent bridge of gas seen in
the channel maps of \citet{sta03} passes just north of our field, and
is misaligned with the optical bar.  This misalignment is shared with
the distributions of very massive main-sequence stars, the brightest red
supergiants, and dust-shrouded protostars--- the effect is perhaps best
seen in the 2MASS starcount maps published by \citet{nik00}.  This
population-dependent bar structure is interpreted as being due to
evolution of the disk structure over time resulting from a combination
of internal and
external perturbations \citep[e.g.][]{dot96}.

Owing to
time-dependent evolution, it is not particularly meaningful to
discuss the ``stellar'' bar of the LMC, because stars of different ages
are distributed differently.  The familiar optical bar is perhaps best
regarded as a ``fossil'' bar, while the less distinct bar traced by
extreme Population~I objects and (possibly) disturbances in the
\ion{H}{1} velocity field \citep[see, e.g.,][]{kim98}
can be thought of as a ``stelliparous''\footnote{From the Latin
{\it stella} = star, $+$ {\it parere} = to bring forth.} bar.
By comparing the morphologies of different types of stars
\citep[e.g.][]{nik00} and star clusters \citep*[e.g.][]{bic92},
we can roughly assign stars more massive than $\approx$6--8
M$_{\sun}$, and clusters of SWB type \citep{sea80} earlier than
III to the stelliparous bar, and older stars to the fossil bar.
This puts the age break between the two systems at roughly
70--200~Myr.  This is intriguingly close to the epoch of the
last major interaction with the Small Magellanic Cloud, an event
which could have dramatically redistributed the angular
momentum of the LMC disk.  Once the bar feature is formed,
it can persist for many orbital periods, with the stars born
and trapped in the bar comprising a dynamical subsystem of the
galaxy \citep[e.g.][]{spa87,she04}.

Throughout the rest of this paper, the term ``bar field'' will
be used to refer to the region shown in Figure~\ref{fig-bar}.
Where a broader context is intended, we will use the terms
fossil bar and stelliparous bar to distinguish these different
morphological systems.  It is important to note that the
bar is historically defined purely on the basis of optical
appearance (Figure~\ref{fig-map}), and the use of the term
does not necessarily imply that a kinematic distinction can
be made between populations with different angular momenta
and energies at the same location.  When population ages
are referred to, we will use the terms very young (0--0.2~Gyr),
young (0.2--1~Gyr), intermediate-age (1--10~Gyr),
and old (10~Gyr).  For stellar populations with velocity
dispersions of 10--20 km s$^{-1}$, 0.2~Gyr is enough time to
diffuse throughout the roughly 2.5~kpc length of the fossil bar;
since our results primarily concern intermediate-age
and old stars, they can be taken as representative of the older
populations of the fossil bar.

\section{\caii\ Spectroscopy}
\label{sec-spec}

The near-infrared \ion{Ca}{2} triplet (CaT) coming
from the (3$^{\mathrm P}$D--4$^{\mathrm P}$D)
transition is an extremely useful set of lines
for the measurement of radial velocities and
metallicities in K giants \citep{arm91}.  The
triplet line strength in old, metal-poor red giants
can be empirically calibrated for metallicity
by removing the influence of surface gravity via
a simple linear equation in V magnitude
\citep*{r97b}.  This empirical calibration
has recently been shown to be applicable to
stars nearly as metal-rich as the Sun and as young as
2.5~Gyr \citep[][paper~III]{col04}.  The empirical
calibration of the CaT to V and [Fe/H] for red giants
is supported by theoretical arguments \citep*{jor92}
as well as by examination of large spectral libraries
\citep{cen02}.  The three triplet lines, at
$\lambda \lambda$ = 8498, 8542, 8662 \AA, are
among the strongest spectral features in K giants,
and fall neatly between regions of strong telluric
H$_2$O absorption.  This has made it an extremely
popular method for the measurement of abundances
in interemediate-age and old stars in dwarf galaxies
throughout the Local Group; the pace of this work
has greatly accelerated with the advent of multiobject
spectrographs and 8--10 meter class telescopes.

\subsection{Target Selection}
\label{sec-targ}

\begin{figure}
%\plotone{AACole.lanl3.eps}
\plotfiddle{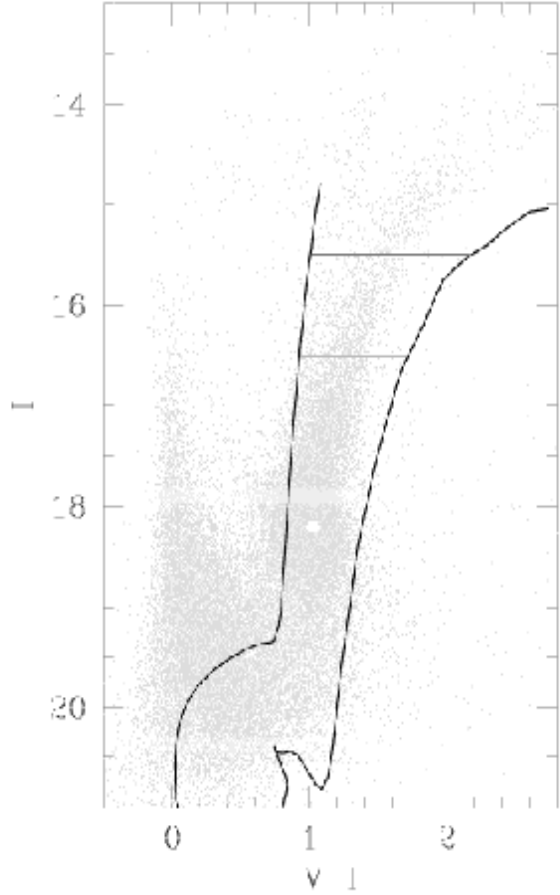}{0in}{0}{288}{396}{108}{288}
\caption{Color-magnitude diagram of the bar
field \citep{sme04}, showing the RGB region between 15.5 $\leq$
I $\leq$ 16.5 from which our targets were selected.  Padua isochrones
for an age of 2.5~Gyr bound the target region in color; the bluer
track has Z = 0.0001, and the redder has Z = 0.019.  The white dot
marks the centroid of the red clump, which will be used in the
metallicity calculation.
\label{fig-sel}}
\end{figure}

We picked our targets to be far enough down the
giant branch that we could avoid M stars while
still fully sampling the color range of the red
giant branch (RGB).  Our color-magnitude diagram
of the region, obtained at CTIO \citep{sme04}
is shown in Figure~\ref{fig-sel}, with our targets
highlighted by heavy points.  We picked our stars
to have 15.5 $\leq$ I $\leq$ 16.5, and to have
V$-$I colors bracketed by two isochrones with
log(t) = 9.40 (age = 2.5~Gyr) from
\citet{gir00}.  The selection region is bounded
on the blue side by the Z = 0.0001 ([Fe/H] = $-$2.3)
isochrone and on the red by the Z = 0.019 ([Fe/H] = 0.0)
isochrone.  The selected targets effectively span
the observed width of the RGB and so the the exact color
limits are not critical to the results.  Some stars
brighter than the I = 15.5 cutoff were observed when
a slit would otherwise have gone unassigned.

Astrometry and photometry for the stars in the central block of
four FORS2 fields (see Figure~\ref{fig-bar}) were taken from
CTIO data published in \citet{sme04}.  These core fields were
the originally-intended spatial extent of our spectroscopic
survey.  However, the southeast quadrant of the core, around
NGC~1950, was unexpectedly crowded with very bright stars and
appeared to have a large young population (possibly in an unbound
corona of cluster stars).  Thus we added three additional flanking
fields around the region with CTIO photometry.  For these
flanking fields, we selected targets based on data
from the OGLE-II survey \citep{uda00}.  Comparison of stars
with measurements from both sources found
\begin{eqnarray*}
I_{CTIO} = I_{OGLE} - 0.007,\;\; \sigma = 0.044\; \mathrm{mag},\\
(V-I)_{CTIO} = (V-I)_{OGLE} + 0.005,\;\; \sigma = 0.050\; \mathrm{mag}.
\end{eqnarray*}
For analysis using optical photometry, we use the CTIO photometry,
or the transformed OGLE-II photometry.

The position of each target was
confirmed using FORS2 preimages obtained 
in service mode several weeks prior to the
observing run, and the slits were assigned using the FORS
Instrument Mask Simulator (FIMS) software distributed by ESO.

\subsection{Data Acquisition \& Reduction}
\label{sec-obs}

The spectroscopic observations were made in Visitor Mode at the
Yepun (VLT-UT4) 8.2-m telescope at ESO's Paranal
Observatory, on the nights of 24--26 December 2002.
We used the FORS2 spectrograph in multi-object (MOS)
mode, with the 1028z$+$29 grism and OG590$+$32 order
blocking filter.   In this configuration, the FORS2
field is covered by a mechanical assembly of 19 slit
jaws, each 20--22 arcseconds long, that can be
arbitrarily positioned along the horizontal
(East-West) axis of the field.  We chose to use a
constant slit-width of 1 arcsecond for ease of calibration.
The spectral images were recorded on two 2k$\times$2k
MIT/LL CCDs, which have a read noise of 2.7 electrons
and an inverse gain of 0.8 $e^-$ ADU$^{-1}$.  The
physical pixels were binned 2$\times$2, yielding
a plate scale of 0.25 arcsec pixel$^{-1}$.  The
resulting spectra cover 1700 \AA, with a central
wavelength near 8500 \AA\ and dispersion 0.85 \AA\
pixel$^{-1}$ (resolution 2--3 \AA).  The FORS2 field
is 6.8 arcmin across, but is limited to 4.8 arcmin
usable width in the dispersion direction in order to
keep important spectral features from falling off the
ends of the CCD.

The log of observations is given in Table~\ref{tab-obs},
which gives the field names and centers, time of observation,
the seeing measured by the differential image motion monitor (DIMM),
and the number
of RGB targets recovered from each setup.  The table includes
the same information for the 12 Galactic star clusters that
were used as metallicity calibrators ({\it q.v.}).
Each LMC setup is identified by a number corresponding to
its position, and a unique suffix of one or more letters
that refers to the slit configuration files produced by
FIMS.  Each configuration was observed twice,
with offsets of 3 arcsec between exposures, to ameliorate
the effects of cosmic rays, bad pixels, and sky fringing.
The total exposure time in each setup was 2$\times$600~sec,
yielding typical signal-to-noise values of S/N $>$30 per pixel.
The seeing varied between 0$\farcs$5 $\leq$ FWHM $\leq$ 1$\farcs$4
during the run, with a median value around 0.8 arcsec.

Calibration exposures were taken in daytime, under the FORS2
Instrument Team's standard calibration plan.  These comprised
lamp flat-fields with two different illumination configurations
and He-Ne-Ar lamp exposures for each slit configuration.  Two
lamp settings are required for the flat-fields because of
parasitic light in the internal FORS2 calibration assembly
(T. Szeifert 2003, private communication).  Owing to the large
number of setups in our program, twilight flats were impractical.
All basic data reduction steps were performed under
IRAF\footnote{IRAF is distributed by the National Optical Astronomy
Observatories, which are operated  by the Association of Universities
for Research in Astronomy, Inc., under cooperative agreement with the
National Science Foundation.}.  We fitted and subtracted the scaled overscan
region, trimmed the image, and divided by the appropriately combined
lamp flats within the {\tt ccdred} package.

Spectroscopic extractions were performed with {\tt hydra}, an
IRAF package for handling multislit spectra.  Our targets were
bright enough that the object trace could be extracted directly
from the science exposures.  Across the $y$-axis of the CCD,
the curvature of the trace along the $x$-direction varied significantly,
but could in all cases be fit with a low-order polynomial.   Because
of the high spectral density and signal-to-noise of night-sky
emission lines--- primarily OH \citep{ost92} and O$_2$ \citep{ost96}---
we used these lines to dispersion correct each spectrum directly
instead of using the arc lamps.  Typically, the 30 or so strongest
emission lines were used in the wavelength solutions, giving a
typical root-mean-square (rms) scatter of 0.04--0.08 \AA.  Because
of the scatter in target positions across the dispersion direction
of the field, individual spectra can reach wavelengths as blue
as 7200 \AA, or as red as 10,100 \AA; most were centered close
to 8500 \AA, covering the approximate range
7600 $\leq$ $\lambda$ $\leq$ 9400 \AA.

\begin{figure}
\plotone{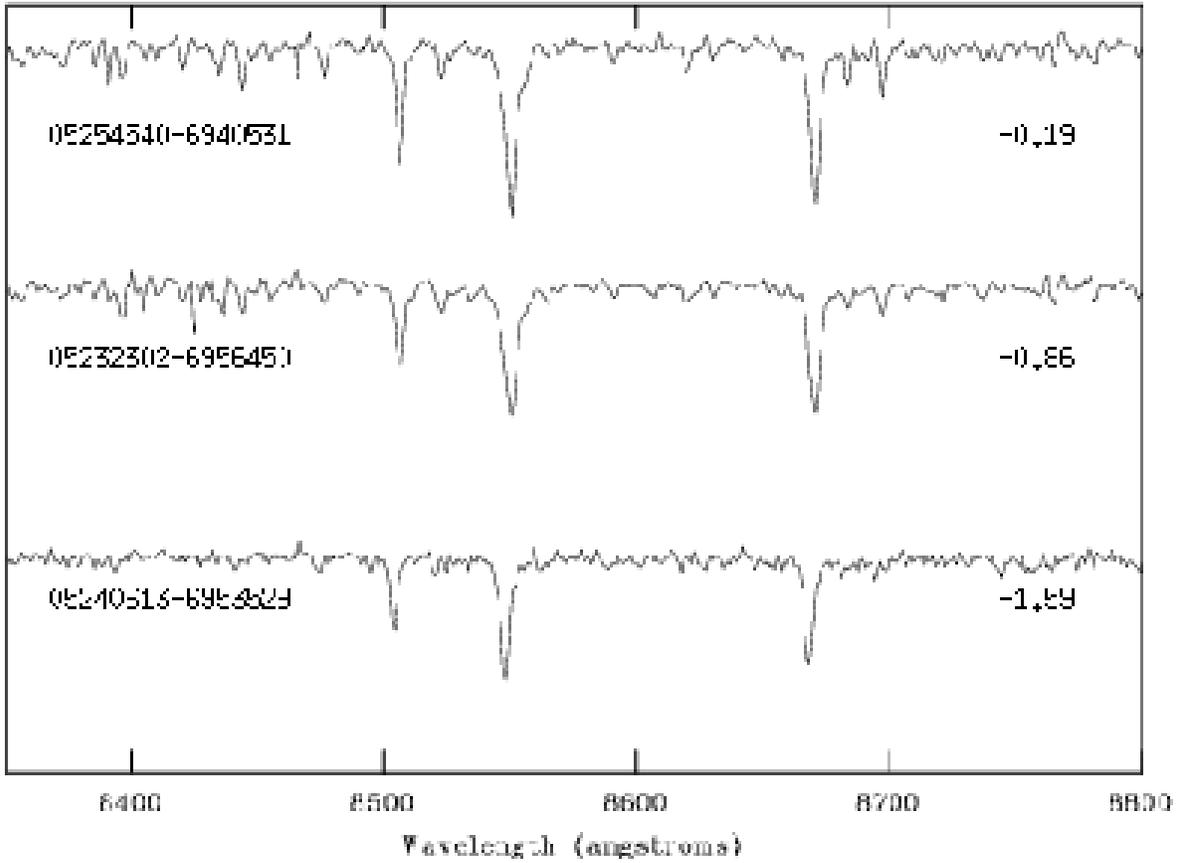}
\caption{Sample spectra of LMC red giants
showing the typical data quality for stars in the fainter half
of our sample.  All three stars have V $\approx$ 17.5; they are
labelled by their 2MASS identifications and metallicities, showing
the change in CaT line strength with [Fe/H].
\label{fig-spec}}
\end{figure}

Extraction to one-dimensional spectra was performed within the
{\tt apall} tasks.  Sky subtraction was achieved using one-dimensional
fits to the background perpendicular to the dispersion direction.  Because
the targets are bright compared to the sky, and the slits are long compared
to the seeing disk, this presented few difficulties.  An exception was when
the stars fell near the ends of the slitlets; in these cases the sky region
was chosen interactively and adjusted to produce the cleanest extracted
object spectrum in the region around the CaT.  Some stars very close to the
top or bottom of the CCD frames showed high sky residuals and were excluded
from subsequent analysis.  The dispersion-corrected spectra were combined
using {\tt scombine} to minimize the effects of bad pixels and cosmic rays.
Each spectrum was continuum normalized by fitting a polynomial to the 
spectrum,
excluding the CaT and regions of strong water vapor absorption.  Sample
spectra are shown in Figure \ref{fig-spec}.

Each extracted RGB star and its FORS2 field identifier are listed
in Table~\ref{tab-dat}, with the VI magnitudes from the sources
listed above.  The stars are identified by their number in the
2MASS point source catalogue \citep{cut03}, except where an
unambiguous identification was not possible; in these cases
the number from the OGLE-II catalogue \citep{uda00} is used.
If a target lies in or around a feature of interest
in Figure~\ref{fig-bar} or has unusual spectral characteristics,
this is noted as well.  The full table is available in the
electronic version of the Journal.

\subsection{Radial Velocities}
\label{sec-vel}

We are interested in stellar radial velocities in order
to reject possible foreground Milky Way stars, and to
search for correlations between the moments of the
velocity distribution and metallicity that could help
distinguish between different stellar populations.
We performed Fourier cross-correlation \citep{ton79}
between our target spectra and the spectra of template
stars of known radial velocity.  24 red giants
in Galactic star clusters were used as templates;
these were a subset of the stars used in our metallicity
calibration (Paper~III), ensuring a good spectral
match between templates and program stars.  We used the IRAF
{\tt fxcor} task to perform the cross-correlation, and
the radial velocities were found from the average of
velocity offsets from each template, weighted by the
random error and the height of the correlation peaks.
The observed velocities were then corrected to the
heliocentric reference frame for subsequent analysis.

Because the stellar image was smaller than the slit
width in most cases, there were in many cases slight
misalignments between the slit centers and the stellar
centroids.  This effect propagates into a potentially
large systematic error in the observed radial velocity
\citep[e.g.][]{irw02}.  We can correct for this velocity
offset if we know the magnitude of the offset in pixels
between the centroid of the stellar image and the centerline
of the slit on the CCD.  Images taken through the slit mask,
without the grism, prior to each exposure were used to 
determine this offset.  Each through-slit exposure images
a patch of sky $\approx$21 arcsec long by 1 arcsec wide
onto the CCD for each target.  The stellar centroid is
determined by a simple profile fit to the through-slit
image, while the position of the slit itself is measured
from 1-dimensional fits to the profile of the sky, excluding
the stellar flux.  The typical offset was less than 0.3 pixels,
compared to the slit width of 4 pixels.
When a nonzero offset in the dispersion
direction was found, we applied corrections to the measured
radial velocities based on the dispersion solution measured
from the night-sky emission lines.  
With our spectral
resolution of $\approx$29.5~km~s$^{-1}$~pixel$^{-1}$,
the resulting velocity corrections ranged from
$| \Delta v|$ = 0 to 32~km~s$^{-1}$, with a mean
correction of $-$0.05 km~s$^{-1}$, and a mean
absolute correction of 8.5 km~s$^{-1}$.  We estimate
that our centroiding accuracy is roughly a quarter
of a pixel, or $\approx$7~km~s$^{-1}$, and we therefore
add this in quadrature to the error in the cross-correlation
for our final error estimates.  The heliocentric velocities
and their associated errors are given in Table~\ref{tab-dat}.

The mean radial velocity of our sample is
V$_{\sun}$ = 257~km~s$^{-1}$, with a root-mean-square
dispersion of 25~km~s$^{-1}$ about the mean.
We found no stars with velocities characteristic
of the Milky Way disk (V$_{\sun}$ $\la$ 100 km~s~$^{-1}$),
and the observed velocity range of
174 km~s$^{-1}$ $\leq$ V$_{\sun}$ $\leq$ 336 km~s$^{-1}$
is entirely consistent with the known range of LMC radial
velocities \citep[e.g.][]{zha03}.  Some Galactic halo
giants have similar velocities, but since they are far
fewer in number than disk stars, the contamination rate
is negligible.  The histogram of heliocentric
radial velocities is shown in Figure~\ref{fig-vel}.
For comparison to the expected distribution,
a thin disk model for the velocity is overplotted:
the mean is derived from the equations in \citet{van02}
to be 260 km~s$^{-1}$, and the dispersion of 24 km~s$^{-1}$
is taken from \citet{zha03}.

\subsection{Equivalent Widths and Abundances}
\label{sec-eqw}

We used the program {\tt ew},
described in Paper~III, to measure
the equivalent widths of the CaT lines by fitting each
of the three lines by the sum of a Gaussian plus a
Lorentzian, constrained to have a common line center.
This was deemed necessary to account for the very
strong damping wings of the lines.  The profile fits
were integrated over the line bandpasses \citep{arm88}
to yield the pseudo-equivalent widths.  Error estimates
were obtained by measuring the root-mean-square scatter
of the data about the profile fits.  The summed equivalent
widths of the three lines ranged from
3.5 \AA\ $\leq$ $\Sigma$W $\leq$ 10 \AA, with typical
errors of $\approx$2\%.  Table~\ref{tab-dat} gives these
values for each target.

Because the relation between CaT equivalent width and metallicity
is empirically defined, and because there have been hints that
the calibration becomes nonlinear at the high-metallicity end
\citep[e.g.][]{car01}, we observed red giants in 12 Galactic
star clusters to define the relation.
The clusters span the metallicity range $-$2.0 $\leq$ [Fe/H]
$\leq$ $-$0.1, and the age range 2.5~Gyr $\leq$ age $\leq$ $\sim$12~Gyr
(see Paper~III for details).
While many of the bar red giants are probably younger than
2.5~Gyr (q.v.), extrapolation of the calibration to ages $\la$1~Gyr
does not seem unreasonable \citep{cen02}.  The calibration
relies on the empirical fact that red giants of a single
metallicity follow a linear relation between $\Sigma$W and
their V magnitude above the horizontal branch, (V$-$V$_{\mathrm{HB}}$):\\
\begin{equation}
W^{\prime}([Fe/H]) \equiv \Sigma W + 0.73 (V - V_{HB}),\\
\end{equation}
which then leads to the following relation between the
reduced equivalent width, W$^{\prime}$, and [Fe/H]:\\
\begin{equation}
[Fe/H] = (-2.966 \pm0.032) + (0.362 \pm0.014) W^{\prime},\\
\end{equation}
with rms scatter $\sigma$ = 0.07 dex.  The distribution of
target stars in the (V$-$V$_{\mathrm{HB}}$), $\Sigma$W
plane is shown in Figure~\ref{fig-cal}, with isometallicity
lines shown for reference.  When comparing to
abundances of Galactic star clusters, it is important to
remember that this calibration is derived with respect to
the abundance scale derived by \citet[][CG97]{car97} for globular
clusters, and the compilation of \citet{fri02} for open
clusters.  The globular cluster and open cluster abundance scales
are thought to be consistent; work is in progress to obtain
a homogeneous set of calcium abundances from high-dispersion
spectroscopy for a large sample of clusters so that in the
future measurements can be calibrated to a single system
\citep{bos04}.

To derive [Fe/H], we adopt the horizontal branch
magnitude V$_{\mathrm{HB}}$ = 19.22, based on
our WFPC2 and CTIO photometry (Paper~II)
Morphologically, this feature is really a red clump
and not a horizontal branch in the strict sense
(see Figure~\ref{fig-sel}); as shown in Paper~III, this
does not affect our metallicity determinations.
The red clump has a V magnitude dispersion of $\pm$0.12
mag, which we propagate through into our metallicity
error estimates.  The total random 1$\sigma$ error on each
metallicity measurement is 0.1--0.2 dex, with an average
value of $\pm$0.14.  The derived metallicities
and their estimated 1$\sigma$ errors are given
in Table~\ref{tab-feh}.  The mean of the sample
is [Fe/H] = $-$0.45, with a dispersion of $\pm$0.31
dex.  However, because of the long tail of metal-poor
stars, the median is a better statistical estimator
of the of the typical metallicity, which is
[Fe/H] = $-$0.40.  The interquartile range is
$-$0.51 to $-$0.28, and the 10th and 90th percentiles
of the distribution are, respectively, $-$0.70 and
$-$0.20.  The distribution is plotted in Figure~\ref{fig-mdf}.

\subsection{Derivation of Stellar Ages}
\label{sec-age}

An expanded view of the RGB region of the CMD
is shown in Figure~\ref{fig-rgbz}.
Spectroscopically
observed stars are color-coded by metallicity,
with the ranges chosen for clarity of display.
It is easily seen that
the most metal-poor and metal-rich stars
roughly divide themselves in color but stars near
the peak of the metallicity distribution, between
$-$0.6 $\leq$ [Fe/H] $\leq$ $-$0.3, span the full
color width of the RGB.  There is even some overlap
between stars with [Fe/H] $\approx$ $-$0.2 and
$\approx$ $-$0.8.  This is a vivid demonstration
that where a large age range is present in a stellar
population, the mean color of the RGB is an innacurate
measure of the metallicity.  Some of our faintest,
reddest stars are far more metal-poor than would be
expected, and could be differentially reddened.

There are two additional
points to take from Figure~\ref{fig-rgbz}:  first, that
although we searched well to the blue of the bulk
of the RGB based on the expectation that metal-poor
red giants should be bluer than their metal-rich
counterparts, most of the bluest RGB stars are
in fact relatively metal-rich; and second, the
broad color range of stars at the peak of the
MDF is indicative of an extremely large range
of stellar ages accompanied by little chemical enrichment
over time.  This encourages us to quantify the
age distribution and age-metallicity relation.

We adopt the procedure described in Paper~II
to derive isochrone ages for each of our target
stars.  We use a program developed by one
of us (AAC) to place isochrones of arbitrary
metallicity in the color-magnitude diagram,
and linearly interpolate in the logarithm
of the age to find the ([Fe/H], log(t)) pair
that reproduces the stellar metallicity and
location in the CMD.
This is not a precise technique, because
the effective temperature of a red giant is
principally controlled by its convective envelope
opacity, which is largely a function of the
abundance of heavy elements in the star
\citep[e.g.,][]{hoy55}.  However, the common
perception that the temperature of a red giant
is largely independent of its mass is inaccurate.

\citet*{hay62} demonstrated that because red giants
are almost entirely convective in their interior
they share a common envelope structure.  Using
analytic homology relations, they showed that
when radiation pressure is unimportant, red giants
must evolve approximately along tracks described
by a relation of the form\\
\begin{equation}
M^{\frac{1}{2}}\, R^{\frac{3}{2}} = \mathrm{constant}.\\
\end{equation}
Using\\
\begin{equation}
R^2 = (\frac{L}{4\pi \sigma T_{eff}^4})\\
\end{equation}
we derive that, insofar as the conditions in
the helium core have only a small effect on the
outer envelope of the red giant,\\
\begin{equation}
M^{\frac{1}{2}}\, L^{\frac{3}{4}}\, T_{eff}^{-3} = \mathrm{constant}.\\
\end{equation}
Thus from basic physical considerations, we
expect that for constant luminosity
T$_{\mathrm{eff}}$ $\propto$ M$^{\frac{1}{6}}$;
this is similar to the dependence discovered in numerical
models by \citet{swe78} and recovered in modern
isochrone sets \citep[e.g.,][]{gir00}.  The magnitude
of dT$_{\mathrm{eff}}$/dM varies with the mass and
luminosity of the giant, but is of order
300--500 K/M$_{\sun}$.  In the context of Figure~8,
this means that two stars of the same abundance and
magnitude will have the same color if they have the
same mass (and hence, age).  If one star is more 
massive (younger) than another, it will be bluer.
This color difference is translated into an age difference
using the published isochrones.

Given representative
values of d(V$-$I)/dT$_{\mathrm{eff}}$
\citep*[e.g.,][]{bes98} and the (strongly decreasing)
function dM/dt, we find that for perfectly known
metallicity and distance, V$-$I photometry good to
2\% could be sufficient to provide an age with $\sim$10\%
accuracy.  In practice, random errors
in the metallicity measurement dominate the uncertainty,
and our typical age errors are of order 60--100\%
(0.2--0.3 in the log).  It is important not to
assign undue weight to the age estimate of an
individual star, but to use large samples of
stars to beat down the random error and thereby
glean some information about the mean age-metallicity
relation.  We tested our technique on a small sample
of star clusters in Paper~II, and found reasonable
agreement with main-sequence turnoff ages, albeit with
large scatter.  Studies of dwarf spheroidal galaxies
\citep[e.g.][]{tol01,bos04b} are generally supportive
of the idea that {\it some} age information can be
gleaned from a combination of accurate RGB metallicities
and colors.  In the most detailed published study of 
four dwarf spheroidals \citep{tol03}, the derived age
distributions showed broad agreement with the star-formation
histories derived from main-sequence turnoff photometry.
However, a significant decrease in sensitivity was noted
for abundances much below [Fe/H] $\lesssim$ $-$1.  An
in-progress study
of fifty-eight red giants in the Carina dwarf spheroidal by one
of us (TSH) will make an extremely interesting test case;
first indications are that the most metal-rich stars
([Fe/H] $>$ $-$1.5) are on average a factor of four younger
than the stars with [Fe/H] $<$ $-$2.   However, finite
errors on the observed metallicities \citep[see][]{sme99b}
make it unlikely that RGB-derived ages will ever clearly
resolve the discrete epochs of star-formation
derived from main-sequence turnoff photometry as in 
\citet*{hur98}.

A potentially major contributor to the
error budget is the uncertainty in
relative abundances of the various
elements heavier than helium.
Because the LMC has experienced a different
chemical evolution history than the Milky Way,
the scaled-Solar abudance
ratios cannot be assumed to apply.  Evidence
for changing values of [O/Fe] with [Fe/H] in
LMC field red giants has been presented by \citet{smi02}
and the amount of data is increasing rapidly
(Hill et al., in preparation).  \citet{hil00}
and \citet{joh04} have also measured the
changing abundance of $\alpha$-process
elements relative to iron in several massive
LMC star clusters, finding important differences
between the LMC globulars and the standard
Milky Way Population~II abundance mixture.
Because the $\alpha$ elements are major
electron donors in red giant envelopes,
[$\alpha$/Fe] ratios play a major role
in determining the stellar T$_{\mathrm{eff}}$
\citep*[e.g.,][]{sal93}.  Therefore it is
vital in attempting to age date a red giant
with isochrones that the correct relative
abundance blend is used.  For practical
purposes, we represent all deviations from
the scaled-Solar abundance mixture using the
parameter [$\alpha$/Fe], and adopt the
approximate relation between the overall
abundance of electron donors, [M/H], and
[$\alpha$/Fe] given by \citet{sal93}:
\[
\mathrm{[M/H]} \approx \mathrm{[Fe/H]} +
    \log (0.638\; 10^{[\alpha/Fe]}\; + 0.362).\\
\]

Because of the limited amount of data available,
we make the simplifying assumption that
[$\alpha$/Fe] = [O/Fe].
Combining the data from \citet{smi02} and the
preliminary results from Hill et al.\ (in preparation),
we approximate the trend of [O/Fe] with [Fe/H]
by a bilinear relation:
\[ [\alpha/\mathrm{Fe}] = \left\{ \begin{array}
                 {r@{\quad:\quad}l}
0.05 - 0.10\, [\mathrm{Fe/H}] & [\mathrm{Fe/H}] \leq -1 \\
-0.413 - 0.563\, [\mathrm{Fe/H}] & [\mathrm{Fe/H}] > -1
\end{array} \right.   \]
There seems to be scatter of 0.1--0.2 dex about
the mean [O/Fe] at given [Fe/H], but this is
not definitely larger than the measurement uncertainty.
The values of [$\alpha$/Fe] adopted for purposes of
the age calculation are included in
Table~\ref{tab-feh}.

We adopt the LMC ``standard'' distance of
50.1~kpc, based on the distance modulus adopted
by the HST Key Project to determine H$_0$
\citep{fre01}; adopting the LMC disk structure
from \citet{van01}, the bar field is 0.1~kpc
more distant than the LMC center of mass,
giving a distance modulus (m$-$M)$_0$(bar) = 18.50. This
value is slightly higher than (but in good
agreement with) more recent distance determinations
based on eclipsing variables \citep{fit03} and
RR~Lyrae stars \citep{alc04}.  Because of this
good agreement, we adopt an errorbar of $\pm$0.06
mag in the distance modulus.  We
adopt a reddening value E(B$-$V) = 0.06 $\pm$0.03 based
on the discussion in \citet{sta03}
\citep*[also see][]{col02,ski02}.

The Padua isochrones include stars as old
as log(Age/Gyr) = 10.25 (Age = 17.8~Gyr).
These were calculated in order to match the
horizontal branch morphology of the oldest
globular clusters, given what was known
about their distances and the parameterization
for stellar mass loss that went into the models.
There is now very strong evidence from the 
first year of data from the WMAP satellite
that the Universe is 13.7~Gyr old \citep{spe03}.
In the meantime, both revisions in the cluster
distance scale \citep[e.g.][]{rei99} and
updated stellar interior calculations
\citep[e.g.][]{vdb02} have produced a strong
expectation that the oldest globular clusters
are no more than $\approx$13.5~Gyr old. 
While these argue for an age bias in the Padua
isochrones at the old end of the scale, there
is much support for their accuracy at 
intermediate ages.  Studies of clusters
\citep*{bon04,sal04} and the field \citep*{bin00}
have derived ages for the oldest open clusters
and the oldest stars in the Solar neighborhood
in the range of 9--11~Gyr.  These are in 
excellent agreement with independent
measurements of the age of the Milky Way disk:
$\approx$9~Gyr 
from thorium cosmochronometry \citep*{mor92}
and from the faint end of the white dwarf
luminosity function \citep*{leg98,han03}.
To bring the results for our oldest stars
into agreement with the known age of the 
Universe, while at the same time preserving
the success of isochrone measurements of
the age of the Galactic disk, we adopt the
expedient of simply rescaling any ages 
older than 10~Gyr.  In the absence of 
detailed information, we use a linear
function
\[
\mathrm{Age} = 10 + 0.41\;
(\mathrm{Age_{raw}}\; - 10)\quad 
\mathrm{(Gyr)}.\\
\]
The average age shift for the ten stars
affected is $-$1.3~Gyr, well within the 
uncertainty in absolute age-dating of any old stellar 
population, and negligible compared to the 
measurement error in our method.  Because
we adopt broad age bins in our analysis of
the age-metallicity relation, the exact
prescription for enforcing consistency between
isochrone ages and the age of the Universe
has very little effect on our results.

There is ambiguity in the
derived ages, because the evolutionary status
of a star of given L, T$_{\mathrm{eff}}$, [Fe/H],
and [$\alpha$/Fe] is not known {\it a priori}.
We have assumed that all stars are first-ascent
red giants, except where the isochrones invalidate
the assumption; in these cases the age has been
derived assuming the star in on the asymptotic
giant branch.  If an assumed RGB star is in fact
on the AGB, our derived age will be roughly 30\%
too young.
The derived age estimates and random
errors, expressed in logarithmic scale, are
given in Table \ref{tab-feh}.  There is a
concentration of stars at 13.7~Gyr caused
by eleven stars that were too red for the oldest
isochrone, the ages of which were set equal
to the age of the Universe (log A = 10.13).
For our choice
of isochrone set, reddening, and [$\alpha$/Fe]
ratios, the median age of RGB stars in the bar
field is 2~Gyr.  The interquartile age range is
1.4--3.4~Gyr, and 90\% of the RGB stars in this field
are younger than 6~Gyr.

\section{Interpretation \& Analysis}
\label{sec-int}

\subsection{The Metallicity Distribution Function}
\label{sec-mdf}

The bar field MDF (Fig.\ \ref{fig-mdf}) is 
a basic datum that should provide strong
new constraints on the inferred history of the LMC
bar based on color-magnitude diagram or spectral
synthesis studies.
The mean and dispersion of
[Fe/H] = $-$0.45 (systematic error $\approx$ $\pm$0.1 dex),
$\sigma$ = 0.31 are not very meaningful statistical
descriptors of the data, owing to the strong asymmetry
of the distribution.  A maximum-likelihood analysis
was used to fit two Gaussian distributions to the
unbinned data; the resulting curve is plotted over
the histogram in Figure~\ref{fig-mdf}.  The curve
is split into a narrow, metal-rich distribution
containing 89\% of the stars, with the remainder
in a broad, metal-poor distribution.  The major
population is described by mean $\mu _1$ = $-$0.37 and
$\sigma _1$ = 0.15, and the minor component by
$\mu _2$ = $-$1.08 and $\sigma _2$ = 0.46.
$\sigma _1$ is barely larger than
our measurement error, suggesting either that we have
been too conservative in our error estimates or that the
intrinsic astrophysical spread in metallicity is
less than $\pm$0.05 dex for this component.

The fraction of metal-poor stars is much smaller
than in the abundance distribution
of long-lived main-sequence stars in the Solar
neighborhood \citep[e.g.,][]{kot02}.  However,
there are strong systematic effects due to
the RGB lifetime that make this comparison
inappropriate.
The bar field red giant MDF is better
compared to the Solar neighborhood data for
G and K type giants brighter than V $\approx$ 5.5
obtained by \citet{mcw90}.  The two distributions
are shown in Figure~\ref{fig-mcw}.  The Solar neighborhood
MDF shows a similar narrow peak at high metallicity,
but is even more deficient in stars with [Fe/H] $\leq$ $-$0.8.
The peak of the McWilliam ({\it metal-rich bar})
sample is at [Fe/H] = $-$0.17
({\it $-$0.37}) and the dispersion
is $\sigma$ = 0.16 ({\it 0.15}) dex.

We can make a direct comparison to the abundance
distribution of the LMC cluster system taken
from the OSSH paper.  OSSH give abundance measurements
for 70 clusters located across the body of the LMC,
taken from similar spectra to those obtained here.
Their equivalent widths are calibrated to globular
and open clusters on a metallicity scale different from the one
we use (Paper~III).  By making a least-squares fit
to the calibrating cluster abundances, we adopt
an {\it estimated} cluster abundance scale
\begin{equation}
\label{eqn-cal}
\mathrm{[Fe/H]_{clus}} \approx -0.212 +
   0.498\, \mathrm{[Fe/H]_{OSSH}} - 0.128\,
   \mathrm{[Fe/H]^2_{OSSH}}.\\
\end{equation}
This recalibration explicitly includes the open
clusters M67 and Melotte~66, and so
supersedes that presented
in Paper~II, which was taken from CG97,
with ad hoc modifications above [Fe/H] = $-$0.5.
Note that some recent measurements
of a subset of LMC clusters at high resolution
and signal-to-noise support a recalibration of the OSSH
measurements, while others support the original OSSH
results \citep[e.g.,][]{hil00,joh04}.  These recent
studies primarily concern clusters with [Fe/H] $\lesssim$ $-$0.9.

The histogram of recalibrated cluster abundances
is plotted with our bar field data in Figure~\ref{fig-o91}.
The cluster MDF is distinctly bimodal, which presents
a visual contrast with the long metal-poor tail of the
bar field MDF.
The main peak of the cluster MDF, corresponding to
the $\approx$1--3~Gyr old clusters, matches up
well with the peak of the bar field MDF, but may
be more asymmetric towards lower metallicities.
Just such a relationship would be predicted by
a model of star formation in which cluster
formation events are shorter and more intense
than field star formation episodes \citep[e.g,][]{bek04}.  However,
we caution against overinterpreting the comparison
in Figure~\ref{fig-o91}, because of the uncertainties
introduced by the different metallicity scales.  

Two additional factors are the accelerated pace of stellar
evolution at low metallicity, and the decrease with
increasing age of the mass range sampled by our RGB
selection region.  This effect (discussed in detail
in Paper~II) effectively biases us against detection
of the older and more metal-poor stars in the field.
This must be taken into consideration when comparing
the cluster and field star MDFs.  For example, 5\% of
the field RGB stars in the bar have metallicities in 
the range $-$1.5 $\leq$ [Fe/H] $\leq$ $-$0.9; the same
fraction of clusters (4 out of 70, or 6 $\pm$3\%)
fall in this range.
Because of the bias against metal-poor field giants,
the true relative fraction of astrated mass in this
metallicity range is likely to be some 2--3 times higher,
erasing any suggestion that the bar field MDF has a bimodality
similar to the cluster MDF.
This effect should apply even more strongly to the most 
metal-poor (oldest) field stars, bringing the {\it observed}
fraction of metal-poor stars with [Fe/H] $<$ $-$1.5 (4\%)
approximately into line with the observed cluster fraction
below $-$1.5, that is, 8 out of 70 (11 $\pm$4\%).  Because
of the strong role of stellar age in determining the 
number of RGB stars in our selection window per unit 
stellar mass created, a detailed comparison of the field star
and cluster metallicity distributions must await a joint
analysis of the color-magnitude diagram and metallicity 
distribution together.

Independently of metallicity scales and sample biases,
the bar field MDF has a different shape (unimodal with
tail) than does the cluster sample (bimodal with slight overlap).  This
confirms the trend found in the inner disk in Paper~II,
and extends it into the center of the LMC.   We find that
the bar field is closer in shape to the cluster MDF than
the inner disk samples in Paper~II.  For example, the
fraction
of field stars falling into the cluster metallicity gap
is smaller in the bar field (5\%) than in Disk~1 (13\%)
or Disk~2 (11\%),
despite the bluer color extent of the sample selection
region in the current study.  This is probably indicative
of the higher fraction of intermediate-age stars in the
bar compared to the disk, expected on both observational
\citep{sme02} and theoretical \citep{bek04} grounds.

The mean metallicity of our sample of
bar field red giants is [Fe/H]$_B$ = $-$0.45, essentially
indistinguishable from that of the D2 field,
[Fe/H]$_{D2}$ = $-$0.46.  Both are more than 0.1 dex
more metal-rich than the D1 field that has
[Fe/H]$_{D1}$ = $-$0.59.  The offset is many times
the formal random error and is highly significant.
All three fields have similar dispersions about the
mean metallicity, $\sigma _{[Fe/H]}$ = 0.32 $\pm$0.01.
The difference between D1 and the other fields
may be related to the location of the
D2 field at the end of the bar and the D1 field in a
region of much lower surface brightness,
far outside the bar-distorted isophotes.  More
data, in widely varying locations, is required before
any firm conclusion about the possibility of spatial
variation in mean metallicity can be reached.
Because of the different selection effects, and
the imminent addition of data from other locations
in the LMC (the Transitional Disk and Eastern fields,
see Figure~\ref{fig-map}), we defer a detailed
comparison of the bar and disk fields to a future
paper.

\subsection{Stellar Kinematics at the LMC Center}
\label{sec-kin}

The radial velocity of our sample is entirely
consistent with the disk rotation curve derived
by \citet{van02} from carbon star velocities,
mostly at projected angular radii greater than
2$\arcdeg$.  Because our field is located almost
directly at the rotation center of the disk, the
rotation signature is expected to be small.  
Therefore we cannot rule out the presence of a
non-rotating (halo or bulge) or slowly-rotating
(thick disk) disk component with these data. 

\begin{figure}
\plotone{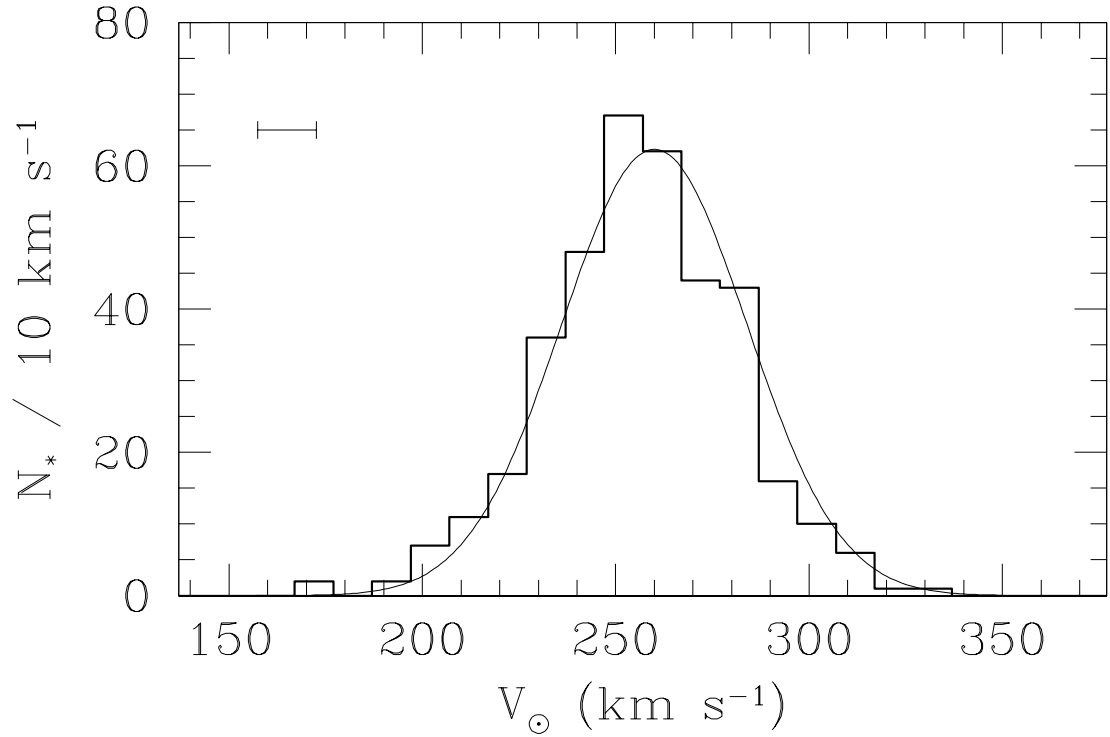}
\caption{Radial velocity histogram of LMC
bar red giants.  The typical 1$\sigma$ velocity errorbar of
$\pm$7.5 km~s$^{-1}$ is shown at upper left.  The smooth curve
is a Gaussian model with mean taken from \citet{van02} and 
standard deviation from \citet{zha03}; it is normalized but
is not otherwise fit to the data.
\label{fig-vel}}
\end{figure}

\begin{figure}
\plotone{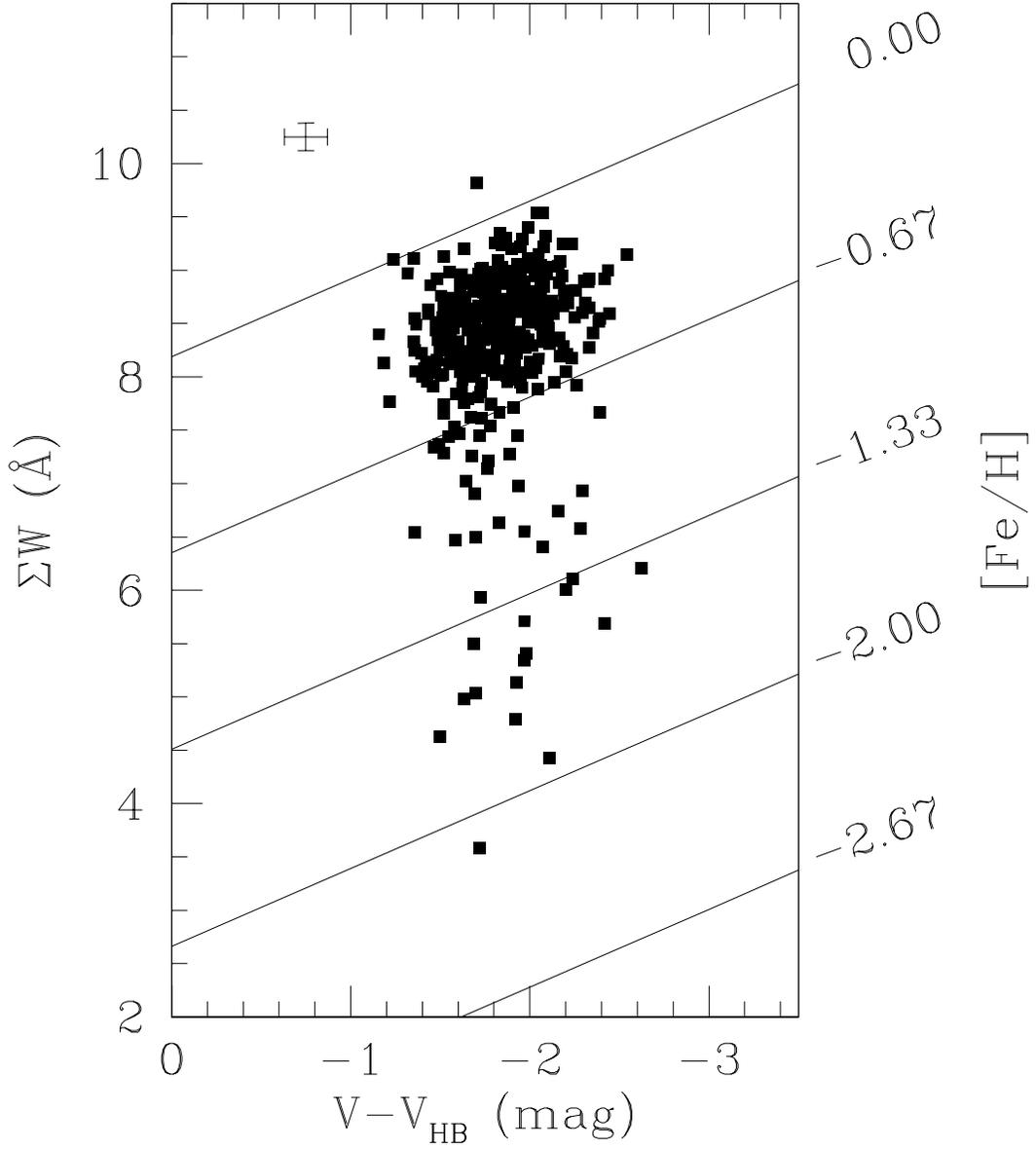}
\caption{Sum of equivalent widths of the 
CaT lines vs.\ V-V$_{\mathrm{HB}}$ for 373 LMC bar red giants.
The typical 1$\sigma$ random error bar is shown at upper left.
The metallicity according to Paper~III is indicated on the scale
at right.
\label{fig-cal}}
\end{figure}

\begin{figure}
\plotone{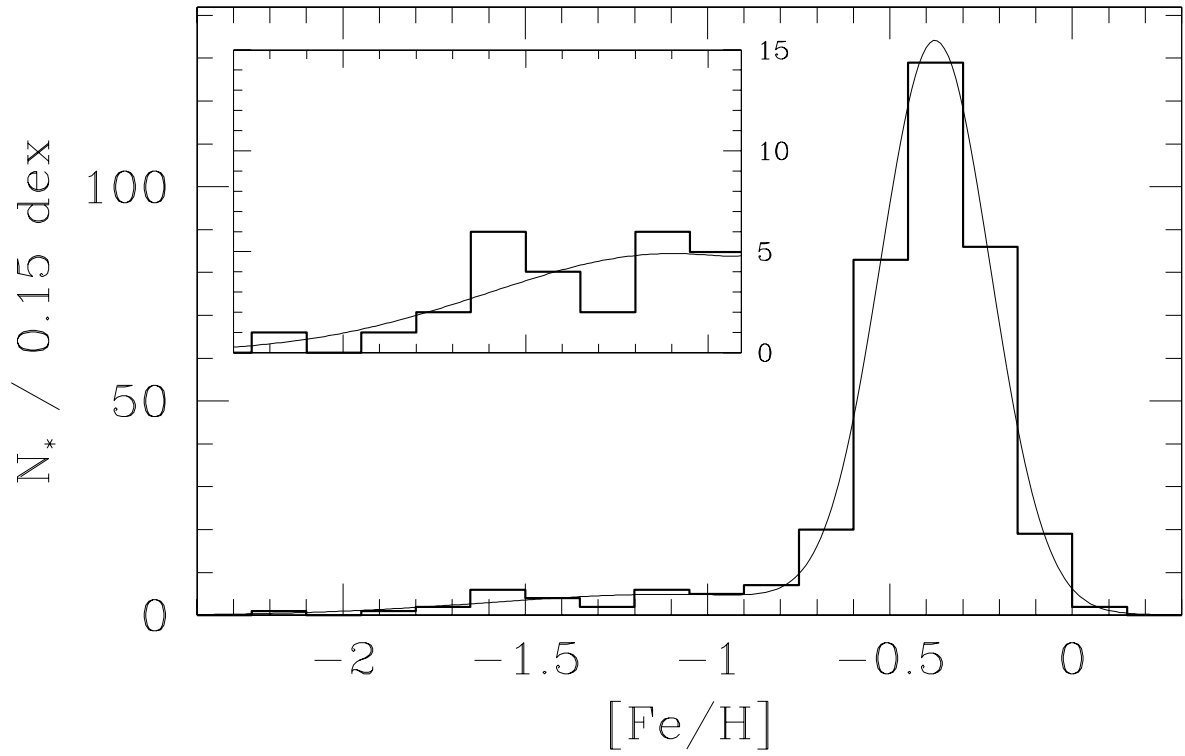}
\caption{Metallicity histogram of the bar
field red giants.  The smooth curve is the sum of two Gaussians
that best match the data.  The inset shows an expanded view of 
the region [Fe/H] $<$ $-$0.9.
\label {fig-mdf}}
\end{figure}

\begin{figure}
\plotone{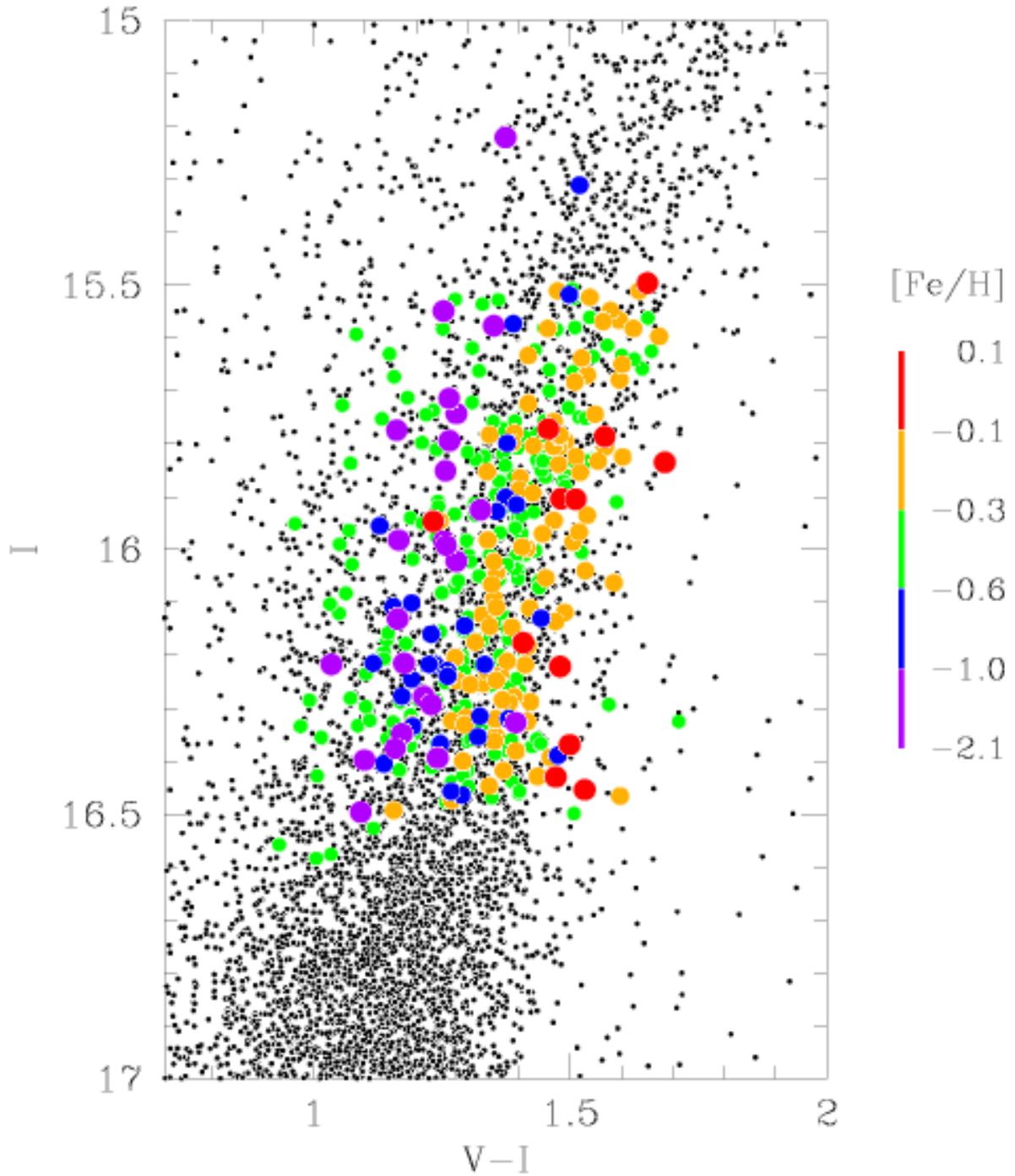}
\caption{CMD of the bar field RGB; the stars
with spectroscopic measurements are color-coded according
to their metallicity.  For the extreme metal-poor
and metal-rich stars the RGB color correlates with
metallicity, but stars near the peak of the MDF
are scattered across the entire width of the RGB.
\label{fig-rgbz}}
\end{figure}

\begin{figure}
\plotone{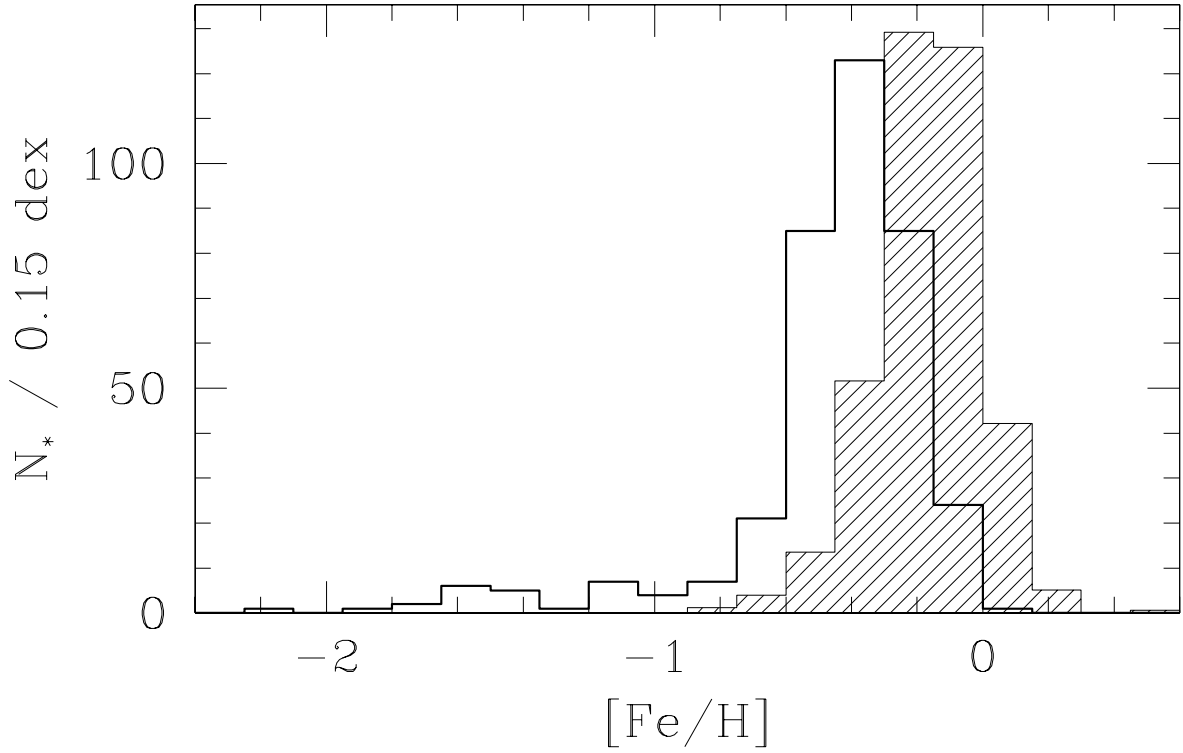}
\caption{MDF of the bar field red giants,
compared to Solar neighborhood data from \citet{mcw90}, normalized
to the same number of stars.  The Solar neighborhood
giants are on average 0.2 dex more metal-rich than the
main population in the LMC bar.
\label{fig-mcw}}
\end{figure}

\begin{figure}
\plotone{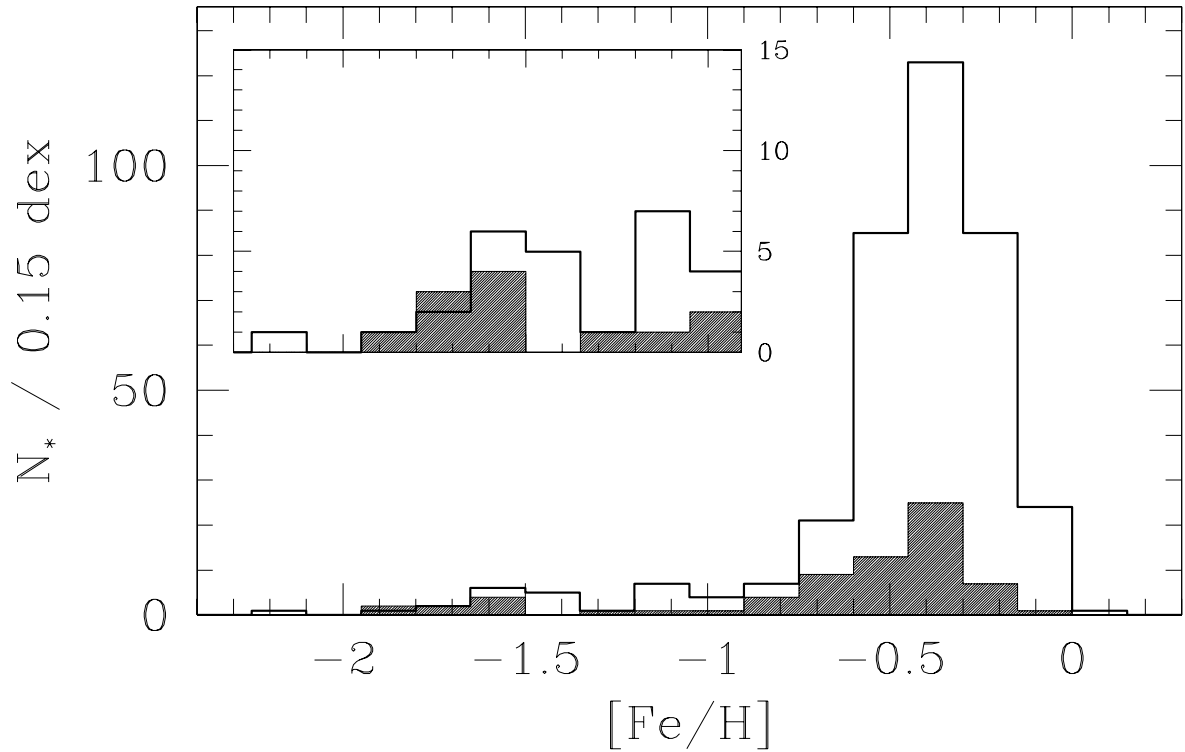}
\caption{MDF of the bar field red
giants (open histogram), compared to 70 LMC star clusters
with abundances from OSSH (shaded histogram).  The OSSH data
have been adjusted to a metallicity scale consistent with
our calibration (see text).  The inset shows an expanded
view of the region [Fe/H] $>$ $-$0.9.
\label{fig-o91}}
\end{figure}

Stars on bar orbits are expected to show large
streaming motions along the long axis of the 
bar \citep[e.g.][]{spa87}.  These could amount
to several tens of kilometers per second, which
would produce a signature in the radial velocity
data as long as the bar does not lie in or nearly
in the plane of the sky.  Detailed predictions
for the kinematic signature of off-center bars
in dwarf galaxies are unavailable, but it seems
likely that stars on bar orbits could be contributing
to the non-Gaussianity in the observed velocity
field (Fig.\ \ref{fig-vel}).  Such non-circular
motions could also be contributing to the
line of sight velocity dispersion, which is 
higher than that predicted by the thin disk
model of \citet{van02}.  

The velocity dispersion along the line of
sight is $\sigma$ = 24.7~$\pm$0.4~km~s$^{-1}$, in excellent
agreement with the general sample of LMC stars measured
by \citet{zha03}.  The dispersion is
slightly higher than the value of 20~km~s$^{-1}$
reported for the global average of LMC carbon stars
by \citet{van02}, and 60\% higher than the
16~km~s$^{-1}$ line
of sight velocity dispersion of \ion{H}{1} gas
reported by \citet{kim98}.  While the measured
dispersion suggests a moderately thick structure,
it is not high enough to imply that the majority
of stars occupy a dynamically hot population such
as a bulge or halo.  It was suggested by \citet{zar04}
that such a structure could account for some 
observations of the morphology and structure of
the inner LMC and give the appearance of a bar.

For many years,
studies of the intermediate-age and old populations
in the LMC have found increased velocity dispersions
with age, up to a limit of roughly 30--35~km~s$^{-1}$
\citep[e.g.][]{hug91,sch92}.  Even the oldest
star clusters appeared to form a thick disk rather
than a spheroid \citep{fre83,sch92}.
By contrast, \citet{min03}
have found a velocity dispersion of 53~km~s$^{-1}$ for
the RR~Lyrae type variables in the area of the bar.
These values roughly bracket what might be generally
expected for a kinematically hot halo in the potential
of the LMC (M $\approx$ 10$^{10}$ M$_{\sun}$).

\begin{figure}
\plotone{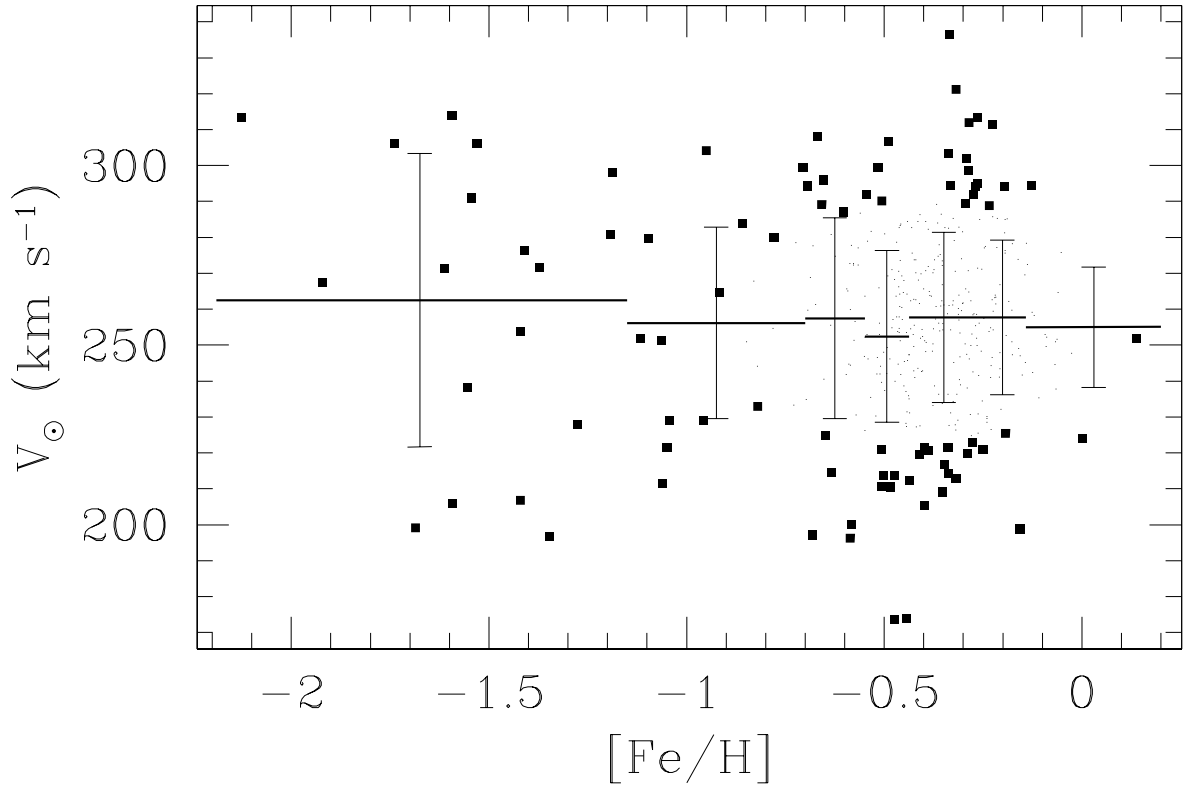}
\caption{Radial velocities of bar field
red giants, plotted against their metallicities.  Small 
symbols are used for clarity where the density of points is
high.  When binned in metallicity, the velocity dispersion
more than doubles between the most metal-rich and metal-poor
stars.
\label{fig-kin}}
\end{figure}

We can test our sample for similar effects by
dividing into several subsamples.  Following the
procedure in Paper~II, we show the line of sight
velocity dispersion of samples in various metallicity
ranges in Table~\ref{tab-kin}.
The plot of radial velocity
vs.\ metallicity is given in Figure~\ref{fig-kin}, showing
that while the metal-rich and metal-poor stars share a
total velocity range of over 100~km~s$^{-1}$, the bulk
of the stars are far more concentrated towards the mean than
the stars more metal-poor than [Fe/H] $\approx$ $-$1.
The mean radial
velocity barely changes with metallicity.
The dispersion starts at
16.7~$\pm$1.6~km~s$^{-1}$ for the most metal-rich
(and presumably youngest) stars, increases dramatically
by the next metallicity bin, and then gradually
grows with decreasing metallicity until the last bin,
when another large jump brings the line of sight
dispersion of the stars below [Fe/H] = $-$1.15
to 40.8~$\pm$1.7~km~s$^{-1}$.  Note that this is
completely in line with the expected line of sight
velocity dispersion for a halo population, but
our sample has neither the size nor the spatial
extent to measure any deviations from a rotating
disk among this minority population.  

\citet{zar04} proposed that the optically-identifed
LMC bar is actually a triaxial bulge.  If we take
its luminosity to be of order $\sim$10$^{8}$ 
L$_{B,\sun}$, then the Faber-Jackson relation would
predict a velocity dispersion in the neighborhood
of $\approx$70~km~s$^{-1}$ for a classical bulge.
Such structures are not associated with late-type
galaxies like the LMC; on the other hand, barred
galaxies are strongly connected with the presence
of box- or peanut-shaped pseudobulges 
\citep[e.g., the very thorough review by][]{kor04}.
The creation of pseudobulges, which are dynamically
colder than classical bulges, is linked to secular
dynamical evolution of the disk and bar.  A pseudobulge
in the LMC would likely have a velocity dispersion of
$\approx$30--40~km~s$^{-1}$, in agreement with what we
observe for our most metal-poor subsample of stars.
However, pseudobulges are created from the general
disk and bar stellar populations, and so there is
no expectation that they should be preferentially
more metal-poor than their surroundings.  Indeed,
\citet{pel96} have found that the stellar populations
of pseudobulges are indistinguishable from those
of the disks in which they are embedded.  This strongly
suggests that the small fraction of stars we observe
at high velocity dispersion and low metallicity does
not owe its existence to the secular heating of the 
disk by the bar.  Irrespective of nomenclature
(``halo'', ``bulge'', or ``pseudobulge'') the
dynamically hot population population in the central
LMC is a very minor contributor to the total stellar 
surface density.

The velocity dispersion we measure is comparable to
our D1 and D2 results from Paper~II.  We do not attempt
to break our bar sample into thin and thick disk components,
because of the unknown influence of the bar on disk
structure and because the distribution in Figure~\ref{fig-vel}
is not particularly well-fit by two Gaussians.  The
probability is that the stellar populations are characterized
by a continuum of velocity dispersions, rising with
age due to gravitational scattering.  It is interesting
to note that the most metal-rich stars are not
significantly hotter than the neutral ISM, suggesting
either that the stellar disk was not strongly heated
by the most recent encounter with the SMC, $\approx$200~Myr
ago, or that the continuing gravitational interactions
with the Milky Way and SMC have kept the \ion{H}{1} from
cooling below this level.

\subsection{The Age-Metallicity Relation}
\label{sec-amr}

\begin{figure}
\plotone{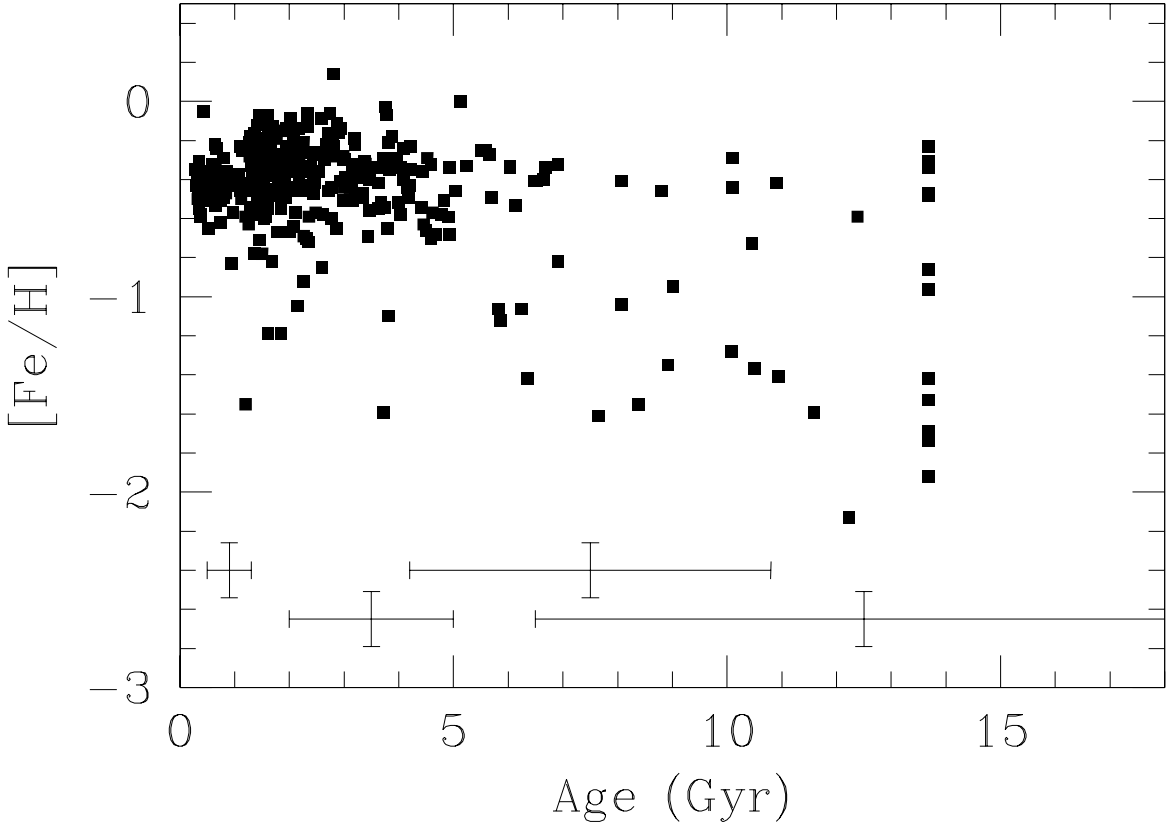}
\caption{Age-metallicity relation for RGB
stars in the bar field.  The ages were derived as described
in the text, with an upper limit of 13.7 Gyr imposed.  The
average 1$\sigma$ error on the age is roughly a factor of two,
as shown by the representative error bars at the bottom of the plot.
\label{fig-amr}}
\end{figure}

\begin{figure}
\plotone{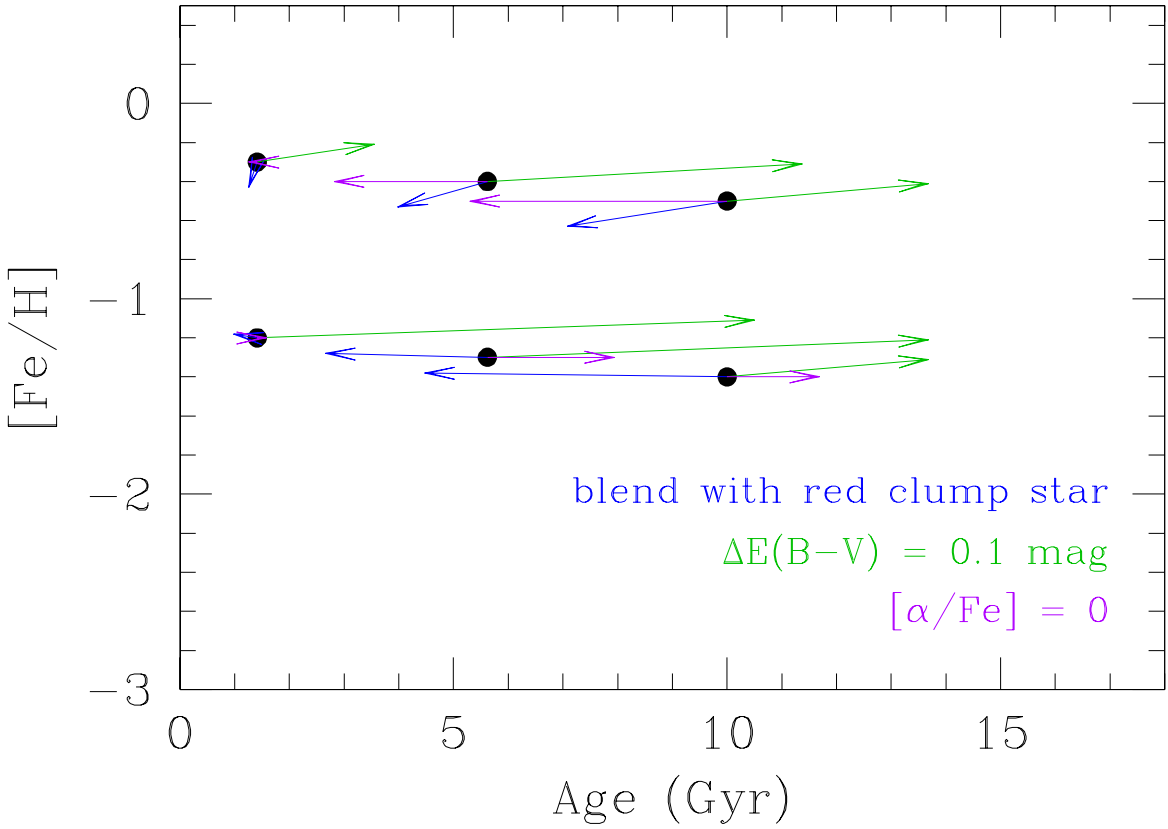}
\caption{Illustration of the systematic 
effects of stellar crowding, differential reddening, and
variable abundance ratios on the age estimates
presented here.  Black circles mark the ``true''
locations of test RGB stars in the age-metallicity
plane, and the arrows show how the measured values
would appear to change as a result of the listed
effects.
\label{fig-sys}}
\end{figure}

The derived age-metallicity plot is shown in
Figure~\ref{fig-amr}.  Typical errorbars are
shown at the bottom of the plot for clarity.
At a given age, there is a large scatter in
metallicity.  Part of this is certainly
a real scatter, and part of it stems from
systematic effects such as differential
reddening or incorrect assumptions about
the [$\alpha$/Fe] ratio.  These effects
are illustrated in Figure~\ref{fig-sys}.
In this diagram the locations of six
``test'' RGB stars with I = 16 have been
plotted in the age-metallicity plane.
Arrows show how the derived ages would change
if the stellar spectra were contaminated with that of a
red clump star of [Fe/H] = $-$0.4 (blue arrows),
if the star was reddened by an additional 0.1
mag in E(B$-$V) (green arrows), or if the
[$\alpha$/Fe] ratio had been assumed to be
Solar instead of following the LMC trend
\citep[e.g.,][]{smi02}.  Note that the
metallicity measurements are robust against
these effects, which primarily affect the
age estimates.

We find 14 stars with [Fe/H] $< -$1 and
age $<$ 10~Gyr.  Similar populations are
much less common in the data for the D1
and D2 fields (Paper~II).  This is not
an artifact of the different selection
criteria adopted: only one of these stars
lies blueward of the adopted edge of the
selection region in the earlier paper.
As we might expect, most of the stars that
the D1 and D2 selection regions would have
missed are quite young, with an average
age below 0.5~Gyr.  The presence of these
stars at the blue side of the RGB indicates
that young stars are much more prevalent in
the bar than in the disk at a radius of one
scale length.
The results of Figure~\ref{fig-sys}
suggest that some fraction of the
apparent intermediate-age, metal-poor stars
may be unresolved blends of genuinely
ancient, metal-poor stars with intermediate-age
red clump stars near the peak of the MDF.
A higher percentage of blended stars in
the bar than in the disk would naturally
be expected because of the much higher
stellar surface density in the bar field.
Some evidence in favor of this interpretation
can be taken from the radial velocities
and age estimates of the two stars that
have [Fe/H] $< -$1 that were flagged during
data reduction as having faint companions
(Table~\ref{tab-dat}).  2MASS point
sources 05225632-6942269 and
05253235-6943137 have ages, respectively,
of 1.9~$\pm$0.8~Gyr and 7.6~$\pm$5~Gyr,
much younger than average for their
metallicities of $-$1.19 and $-$1.61.
Depending on the properties of the
faint companion objects, their true ages
could be much older.

\begin{figure}
\plotone{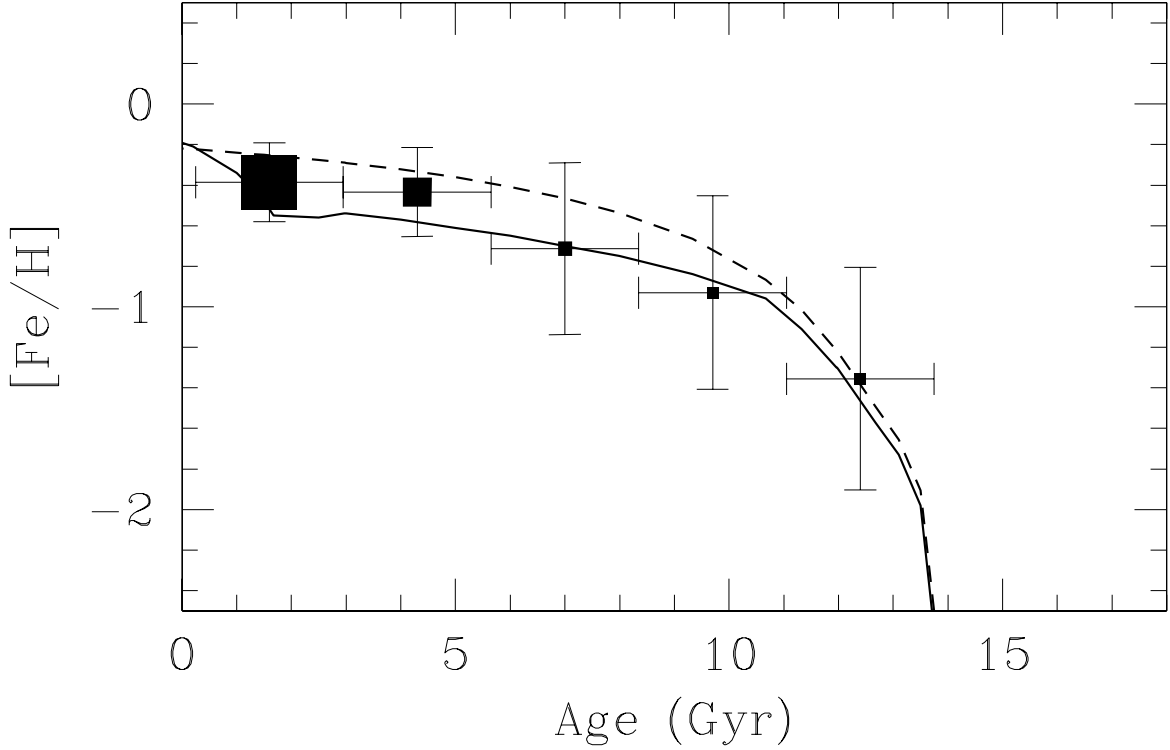}
\caption{Age-metallicity relation for
the bar field, binned into 2.7~Gyr intervals from 0.2--13.7~Gyr.
The symbol size is proportional to the number of stars in each
age bin.  Error bars show the bin width and the rms
dispersion of abundances in each bin.  Five stars suspected
of being differentially reddened have been excluded 
(see text for details).  Chemical evolution models from 
\citet{pag98} have been overplotted, for both bursting 
(solid line) and continuous (dashed line) star-formation
histories.
\label{fig-amrb}}
\end{figure}

Because of the large random errors (up to
a factor of two) and the possibility that
systematic effects may ruin some
individual estimates, the age information
is most usefully interpreted when binned
up to increase signal-to-noise and suppress
the effects of outliers.  We sort the stars
into five equal-age bins 2.7~Gyr wide, and
show the mean metallicity in each bin
in Figure~\ref{fig-amrb}.
The five faintest, reddest stars, measured
perpendicular to the RGB ridgeline (see
Fig.~\ref{fig-rgbz}), have been excluded from
the averaging because differential reddening
is suspected. The vertical errorbars on
each point show the rms scatter about the mean
in each bin, and the horizontal bars denote
the extent of the bin.  The area of each point
is proportional to the number of stars in the
bin, ranging from 10 in the oldest bin to 255
in the youngest.

Figure~\ref{fig-amrb} shows that the metallicity
steadily increases with time: quickly at ancient
times, and then by $\lesssim$0.5 dex over the
past 10~Gyr.  The metallicity scatter {\it appears}
to decrease with time, from $\pm$0.5~dex in the
oldest bin to $\pm$0.2~dex in the youngest.  For
the oldest bin, the likely culprit is the requisite
rapidity of chemical evolution from [Fe/H] $\approx$
$-$3 at the end of the Population~III phase to $\approx$
$-$1 within the first 4~Gyr.  The fact that the
errors in our age estimate are large for the
oldest stars probably introduces additional scatter
by creating mixing between age bins.

The age-metallicity relation we derive is
compared to the results from studies
that have focused on specific sub-populations
of stars in Table~\ref{tab-pops}.  The first
two lines recapitulate the two Gaussian
fits to the MDF derived in section~\ref{sec-mdf}.
Based on the derived ages, the two metallicity
components are split into young and old populations,
although there is obviously considerable overlap
between them.   The remainder of the table
shows the mean metallicities and dispersions
of various stellar populations, arranged by
increasing age.  Where we have been able to
trace the published abundances back to a scale
similar to that used by OSSH, we have applied
Equation~6 to bring the values into line with
our data.  The picture is one of rapid evolution
at early times, followed by a very slight increase
over the past 10~Gyr.

B~dwarfs and Cepheids are
taken to be representative of the very young
stellar populations; their mean abundance is
only very slightly higher than the peak of the
metal-rich component of the field giant MDF
in the bar.  Evidently the chemical evolution
of the LMC has been quite modest over the past
$\sim$10$^9$ years.  Very few tracers
of chemical evolution at intermediate age
have been available to date;
the sample of planetary
nebulae measured by \citet{dop97} contains very few
objects older than the oldest intermediate-age
star clusters.  The only star cluster with an
age of 4--10~Gyr is ESO121-3, with [Fe/H] =
$-$0.93 \citep[OSSH;][]{hil00}.  Although the
AMR appears shallow, we find the mean metallicity
of stars aged 3--6~Gyr to be [Fe/H] = $-$0.46 $\pm$0.02,
compared to $-$0.72 $\pm$0.03 for the stars
aged 6--8~Gyr, nearly a factor of two
difference. 

The low-metallicity
component of our MDF is more metal-rich than the
average of old star clusters or field RR~Lyrae
variables, with a higher dispersion.  This
indicates the continuous nature of field star
formation, in that we have probed a much wider
range of the LMC's history than just the oldest
populations traced by the globular clusters and
RR~Lyraes.  If we just consider the 14 stars estimated
to be older than 10~Gyr and not suspected of
differential reddening,
the mean [Fe/H] = $-$1.31,
with a dispersion of $\pm$0.51.  This is consistent
with the field RR~Lyrae stars of the bar, although
the dispersion is larger than the value of $\pm$0.29
dex in the RR~Lyrae sample of \citet{gra04}.

\subsubsection{Comparison to Models}

Models for the chemical evolution of the LMC
\citep[][(PT98)]{pag98}, based on two different assumed
star formation histories and with the yields
adjusted to fit the ancient globular clusters
and the numerous clusters aged 1--3~Gyr, are
overplotted on our binned age-metallicity relation.
The dashed line shows the chemical evolution
derived from assuming that the star-formation
rate had a broad peak $\approx$10~Gyr ago and
has been very slowly declining since then.  The
solid line marks instead the chemical history
of a model LMC with roughly constant
low level of star formation for most of its
history, that then experienced a factor of six
jump in star formation rate 3~Gyr ago, with
a subsequent rapid decline.
Both smooth and bursting classes of SFH can
reproduce the star cluster age-metallicity
relation, owing to the lack of clusters
between 3--10~Gyr old.

As found in D1 and D2,
the field stars fill in the cluster age gap with
a continuous distribution of ages and metallicities.
This raises the possibility that we can statistically
distinguish between the two cases.  It can be seen
from Figure~\ref{fig-amrb} that stars aged from
$\approx$2--10~Gyr will have the strongest lever
on the models, with little to differentiate between
them at the oldest and youngest times.  For each
observed star, we start from the age derived in
Table~\ref{tab-feh} and calculate the probability
that it was drawn from the age-metallicity relation
appropriate to the bursting or the smooth model,
taking the observational error on [Fe/H] into account.
The relative likelihoods of the models are then
computed by finding the joint probability of
observing the entire ensemble of stars under
each model.  The amount of cosmic scatter
assumed in the model AMR will influence the
results, so we adopt $\sigma _{\mathrm{AMR}}$ = 0.15
as a realistic estimate.

We find that the observed AMR is better
matched by the smooth model from PT98 than their
3-Gyr burst model at the 2$\sigma$ level (95.8\%
confidence).  While the latter is a better
match to the stars aged 5--10~Gyr, these stars
are greatly outnumbered by younger stars, which
have higher abundances than predicted by the
burst model.  However, many
lines of evidence point to a bursting history
of star fomation in the LMC.  \citet{sme02}
have shown that the epoch of increased star formation
rate in our bar field is likely to have occurred
earlier than 3~Gyr ago.  
Figure~\ref{fig-amrb}
makes plain that an earlier burst can be tuned
to match the stars both older and younger than
5~Gyr, and so can be made to be fit the data
better than the smooth model.  Based on the
shape of the AMR predicted by PT98's bursting
model, a burst would be expected to have 
occurred prior to $\approx$5~Gyr ago, but
not much before $\approx$7~Gyr.

Because of the low precision of our age
estimates and the uncertainties in computing
chemical evolution models, a maximum likelihood
calculation of the time and amplitude of a
starburst from these data is unlikely
to produce meaningful astrophysical results.
We defer such an exercise to a future paper,
in which we will simultaneously model the full 
WFPC2 CMDs down
to below the oldest main-sequence turnoff and
the red giant MDF derived here.

\subsection{Spatial Patterns \& Clustering}

We attempted to avoid star clusters and associations
as much as possible, but three stars projected
directly on two clusters (HS 256 and HS 285) crept
into the measured sample.  An additional thirteen
stars (identified in Table~\ref{tab-dat}) in the
close vicinity of five star clusters were also
measured.  Considering the field density and the
apparent blue colors of the clusters on our
preimages, it is unlikely that the stars in the
neighborhoods of the small clusters are bona fide
members.

Neither HS~256 nor HS~285 has a published age
in the literature.  HS~256 partially overlaps
with one of our WFPC2 fields, so it may be
possible to examine its color-magnitude diagram
separately from the field in a future paper.
Two stars of our sample, 2MASS 05223416-6944433
and 05223895-6945007, are seen in projection
against HS~256.  Their radial velocities are
both very low compared to the sample mean,
222 and 227 km~sec$^{-1}$ respectively, which
may indicate their joint membership in a
population with low velocity dispersion
moving towards us at 30 km~sec$^{-1}$ relative
to the mean bar field.  However, the two
stars have very different metallicities:
[Fe/H] = $-$1.05 $\pm$0.12 and $-$0.42 $\pm$0.14.
There is no reason to suspect the quality
of the abundance measurement in either star,
but the widely discrepant values militate against
common cluster membership.   The single star
seen in projection against HS~285 is indistinguishable
from the general field in both its radial velocity
(261 km~sec$^{-1}$) and metallicity ($-$0.37).

We examined maps of the area, searching for
spatial patterns in the radial velocity,
metallicity, and age of the stars.  No strong
evidence for structure in the populations
was observed.  However, there did appear to
be a slight concentration of the reddest
stars into the southwest corner of the field,
near the largest of the dark clouds identified
by Hodge, and some small, chainlike \ion{H}{2}
regions (see Figure~1).  When comparing the
2MASS J$-$K colors of the stars, we found
the reddest stars to be far more evenly dispersed
throughout the bar field.  Because the J$-$K
color is less affected by reddening than is V$-$I,
this is consistent with our interpretation that
the concentration of stars with high V$-$I is
not the effect of high metallicity or old age.

\section{Summary \& Discussion}

The high surface brightness and extreme crowding
of the central regions of the Large Magellanic
Cloud have challenged observers for decades.
In this paper, we
present the first spectroscopic study of
the abundances, kinematics, and age-metallicity
relation for field red giants in the LMC bar.
Taking advantage of the superb image quality
and efficiency of the FORS2 spectrograph at
VLT-UT4 (Yepun), we obtained spectra of 373
red giants in a 200 square arcminute region
of the central bar that includes several fields
singled out for detailed photometric study
with WFPC2 \citep{hol99,sme02}.

We have
derived abundances on a metallicity scale consistent
with those of CG97 (globular clusters)
and \citet[][open clusters]{fri02}
(Paper~III), with internal accuracy of $\pm$0.14
dex per star.  Radial velocities accurate to
$\pm$7.5~km~s$^{-1}$ were measured by Fourier
cross-correlation of our spectra with
template stars of similar spectral type.
We used isochrones from \citet{gir00} and
assumed non-Solar elemental abundance ratios
based on \citet{smi02} and Hill et al.\
(in preparation) to make age estimates
with random errors of roughly 60\%.

Our main results are:

\begin{enumerate}
\item The mean metallicity of red giants in
the central LMC is [Fe/H] = $-$0.45, with a
diserpsion about the mean of $\pm$0.31.  The
distribution can be described by the sum of
two normally distributed populations in the
ratio of 8:1, with the majority (minority) population
having mean [Fe/H] = $-$0.37 ($-$1.08) and
dispersion $\sigma$ = 0.15 (0.46).  Half
the stars have metallicities in the range
$-$0.51 $\leq$ [Fe/H] $\leq$ $-$0.28;
Only 10\% are more metal-poor than [Fe/H]
= $-$0.7.

\item The mean heliocentric radial velocity
of our sample is 257~km~s$^{-1}$
The observed velocity dispersion of 
24.7 $\pm$0.4~km~s$^{-1}$
is typical of intermediate-age LMC stars.
The velocity dispersion increases with
decreasing metallicity, from 16.7 $\pm$1.6~km~s$^{-1}$
for the most metal-rich 5\% of stars, to
40.8 $\pm$1.7~km~s$^{-1}$ for the 5\% of
most metal-deficient stars ([Fe/H] $< -$1.15).
Over most of the intervening range, the
velocity dispersion is roughly constant around
22--27~km~s$^{-1}$.

\item The median age of the stars is roughly 2~Gyr,
with an interquartile range of 1.4--3.4~Gyr.
90\% of the RGB stars appear to be younger than 6~Gyr.
This distribution does not linearly translate to
the variation in star-formation rate over time because
of strong RGB lifetime effects that bias the observed
age distribution towards young stars.

\item The age-metallicity relation is in excellent
agreement with measurements of the old and young
star clusters and other tracer populations.  For
the first time, we observe the evolution of metallicity
over time through the cluster age gap from 3--10~Gyr
ago. The AMR combined with chemical evolution models
appears to favor an increase in star-formation rate
sometime prior to $\approx$5~Gyr ago.

\end{enumerate}

\subsection{Discussion}

The metal-rich component of the bar field MDF is
similar in width to the MDF of solar neighborhood
red giants as measured by \citet{mcw90}, although
the mean is shifted to lower metallicity by 0.2~dex.
The low-metallicity tail of the MDF appears not
to be present in the solar neighborhood; this is
probably because our ``bird's-eye view'' from above
the LMC penetrates the disk at a steep angle,
including populations regardless of the details
of their vertical distribution.

The behavior of the velocity dispersion with
metallicity is also reminiscent of the Milky
Way disk, in which stars are born with low
velocity dispersion that increases quickly
with time for $\approx$2~Gyr, and then remains
roughly constant with age until 10~Gyr \citep{fre02}.
In the Milky Way, this is taken to be the
signature of the thick disk at old times.
In the LMC, the situation is less clear,
and the possibility cannot be eliminated that
the most metal-poor, oldest stars are distributed
in a spheroidal or halo distribution \citep{min03}.
The suggestion of \citet{zar04} that the apparent
bar may in reality a partially obscured, triaxial
bulge seems disfavored by the observed velocity
dispersion, which is much smaller than would be
expected for a classical bulge.  
The vast majority of red giants
appear to be consistent with a thick disk type
distribution.  A box- or
peanut-shaped pseudobulge, with much lower 
velocity dispersion than an r$^{\frac{1}{4}}$ bulge,
some rotational support,
and stellar populations similar to the surrounding
disk is allowed (although by no means required)
by the observations.  
Because of the very close association between
bars and pseudobulges \citep[e.g.,][]{kor04},
it doesn't seem tenable to invoke such a feature
as an {\it alternative} to a bar; rather if a 
boxy structure is present, it would almost 
certainly be {\it additional} to a bar.

The family resemblance to the Milky Way is
less obvious when it comes to the age-metallicity
relation.  There is no obvious AMR in the
Milky Way thin disk \citep{fri02,fre02}, although
one does appear to exist in the thick disk
\citep{ben04}.  Our LMC bar sample shows a
clear increase in mean metallicity over time, with
most of the increase occurring since $\approx$6~Gyr
ago.  However, there {\it are} a few metal-rich stars
among the apparently old stars of the sample,
as in the Milky Way disk.  As in the Milky Way,
our data imply a cosmic abundance scatter of $\pm$
$\approx$0.15 dex at given age; the appearance
of higher scatter at old ages is attributed to the
rapid pace of enrichment in the youth of the galaxy.
A further point of comparison, the possible existence
of a radial metallicity gradient in the LMC disk,
will be addressed in a future paper.
Until radial velocity and detailed abundance analyses
of sufficient sample size and precision are available,
it will remain uncertain how far parallels between
the Milky Way and LMC disks can be taken.

Attempts to understand the LMC's morphology and star
formation history in terms of its status as a satellite
of the Milky Way have a long history \citep[e.g.,][]{mur80}.
It has long been appreciated that tidal interactions
probably have a leading part in determining the
star formation history of both galaxies \citep{sca87},
as well as the internal structure of the LMC \citep{wei00}.
It is instructive to compare our results to the
predictions of recent gasdynamical N-bdoy simulations
of the interaction between the Milky Way/LMC/SMC
triplet \citep{bek04}.

These models predict that the
first era of strong interaction between the two Clouds
occurred $\approx$6--7~Gyr ago; this resulted in the
tidal capture of the SMC by the LMC, produced the
high surface brightness bar of the LMC, and initiated
an epoch of enhanced star-formation.  This epoch of
intense activity culminated in a violent collision
between the Clouds $\approx$3.6~Gyr ago, creating the
generation of 1--3~Gyr old star clusters and raising
the mean metallicity of the LMC by a factor of $\sim$3
during this time.  The \citet{bek04} simulations therefore
predict that the field stars have a broader age distribution
than the clusters, that the intermediate-age populations
are centrally concentrated to the LMC bar, and that the
metallicity began to increase rapidly between 3--6~Gyr ago.
These predictions are borne out by the picture of the LMC's
history that has been built up in \citet{col02,sme02} and
this paper.

In this picture of a tidal origin for the LMC bar and
intermediate-age clusters, a radial abundance gradient
should exist, because the younger, more metal-rich
populations are centrally concentrated.  \citet{bek04}
also predict that a small but non-negligible fraction
of stars older than the first epoch of strong SMC-LMC
interaction will be scattered into a halo-like distribution
with velocity dispersion $\sigma \approx$40~km~s$^{-1}$,
and with higher metallicity and lower age than the
Milky Way halo.  We will explore these issues, through
direct comparison of the kinematics, MDF and AMR in the bar to the
inner disk (Paper~II) and other outer disk fields
(Cole et al., in preparation), in a future paper.
The next step towards a complete view of the history
of the Large Magellanic Cloud is to combine the
abundance information gathered here with the deep
CMD data we have already obtained with HST to self-consistently
model the star formation history and chemical evolution
of this galaxy.

\acknowledgments

This work is based on observations collected at the European Southern
Observatory, Chile, under proposal number 70.B--0398.  Preimaging
data were taken in service mode, thanks to the efforts of the Paranal
Science Operations Staff.
AAC would like to thank T. Szeifert,
E. Mason, F. Clarke, and P. Gandhi for their support and assistance
during the observing run.   Thanks to I. P\'{e}rez-Mart\'{\i}n for
helpful discussions about bars and pseudobulges.
AAC is supported by a fellowship from
the Netherlands Research School for Astronomy (NOVA).  ET
gratefully acknowledges support from a fellowship of the Royal
Netherlands Academy of Arts and Sciences (KNAW).
Support for TSH's research was provided by the National Science Foundation 
through grant AST-0070895. This
publication makes use of data products from the Two Micron All
Sky Survey, which is a joint project of the University of Massachusetts
and IPAC/Caltech, funded by NASA and the NSF.

{}

\begin{landscape}
\begin{deluxetable}{lcccccl}
\tablewidth{0pt}
\tabletypesize{\small}
\tablecaption{Observing Log \label{tab-obs}}
\tablehead{
\colhead{Field/Setup} &
\colhead{$\alpha$} &
\colhead{$\delta$} &
\colhead{Time observed} &
\colhead{Image FWHM} &
\colhead{Targets} &
\colhead{Comments} \\
\colhead{} &
\colhead{(J2000.0)} &
\colhead{(J2000.0)} &
\colhead{(UT)} &
\colhead{(arcsec)} &
\colhead{} &
\colhead{} }
\startdata
NGC 1904     & 05$^{\mathrm h}$24$\fm$2 & $-$24$^{\circ}$31$\arcmin$ &
    2002-12-25/01:13 & 0.7 & 17 & Calibrator: [Fe/H] = $-$1.37 \\
NGC 104      & 00$^{\mathrm h}$26$\fm$5 & $-$71$^{\circ}$51$\arcmin$ &
    2002-12-25/01:25 & 0.7 &  8 & Calibrator: [Fe/H] = $-$0.70 \\
LMC Bar 1/A  & 05$^{\mathrm h}$23$\fm$0 & $-$69$^{\circ}$53$\arcmin$ &
    2002-12-25/01:37 & 0.6 & 19 &    \\
LMC Bar 1/B  &                          &                            &
    2002-12-25/02:12 & 0.6 & 19 &    \\
LMC Bar 1/M  &                          &                            &
    2002-12-25/02:42 & 0.8 & 16 &    \\
LMC Bar 3/E  & 05$^{\mathrm h}$24$\fm$2 & $-$69$^{\circ}$45$\arcmin$ &
    2002-12-25/03:16 & 0.9 & 19 &    \\
LMC Bar 3/F  &                          &                            &
    2002-12-25/03:46 & 1.0 & 19 &    \\
LMC Bar 3/N  &                          &                            &
    2002-12-25/04:17 & 0.8 & 19 &    \\
Melotte 66   & 07$^{\mathrm h}$26$\fm$5 & $-$47$^{\circ}$41$\arcmin$ &
    2002-12-25/04:48 & 0.7 & 14 & Calibrator: [Fe/H] = $-$0.47 \\
Berkeley 39  & 07$^{\mathrm h}$46$\fm$7 & $-$04$^{\circ}$41$\arcmin$ &
    2002-12-25/05:02 & 0.7 & 10 & Calibrator: [Fe/H] = $-$0.26 \\
Berkeley 20  & 05$^{\mathrm h}$32$\fm$6 & $+$00$^{\circ}$10$\arcmin$ &
    2002-12-25/05:16 & 0.8 &  4 & Calibrator: [Fe/H] = $-$0.61 \\
LMC Bar 6/J  & 05$^{\mathrm h}$25$\fm$9 & $-$69$^{\circ}$40$\arcmin$ &
    2002-12-25/05:40 & 0.8 & 18 &    \\
LMC Bar 6/K  &                          &                            &
    2002-12-25/06:06 & 0.9 & 18 &    \\
LMC Bar 6/L  &                          &                            &
    2002-12-25/06:31 & 0.7 & 18 &    \\
\hline
NGC 1851     & 05$^{\mathrm h}$14$\fm$2 & $-$40$^{\circ}$04$\arcmin$ &
    2002-12-26/00:49 & 0.8 & 15 & Calibrator: [Fe/H] = $-$0.98 \\
LMC Bar 2/C  & 05$^{\mathrm h}$23$\fm$0 & $-$69$^{\circ}$45$\arcmin$ &
    2002-12-26/01:04 & 0.8 & 17 &    \\
LMC Bar 2/D  &                          &                            &
    2002-12-26/01:28 & 0.8 & 19 &    \\
LMC Bar 2/AA &                          &                            &
    2002-12-26/01:54 & 0.7 & 18 &    \\
LMC Bar 2/EE &                          &                            &
    2002-12-26/02:18 & 0.7 & 16 &    \\
NGC 2141     & 06$^{\mathrm h}$03$\fm$0 & $+$10$^{\circ}$30$\arcmin$ &
    2002-12-26/04:15 & 1.2 & 15 & Calibrator: [Fe/H] = $-$0.33 \\
NGC 2682     & 08$^{\mathrm h}$51$\fm$4 & $+$11$^{\circ}$48$\arcmin$ &
    2002-12-26/06:36 & 0.7 &  7 & Calibrator: [Fe/H] = $-$0.15 \\
LMC Bar 5/H  & 05$^{\mathrm h}$22$\fm$7 & $-$69$^{\circ}$38$\arcmin$ &
    2002-12-26/06:51 & 0.9 & 18 &    \\
LMC Bar 5/I  &                          &                            &
    2002-12-26/07:15 & 1.0 & 18 &    \\
LMC Bar 5/BB &                          &                            &
    2002-12-26/07:39 & 0.6 & 18 &    \\
NGC 4590     & 12$^{\mathrm h}$39$\fm$5 & $-$26$^{\circ}$45$\arcmin$ &
    2002-12-26/08:20 & 0.6 &  9 & Calibrator: [Fe/H] = $-$1.99 \\
\hline
LMC Bar 7/O  & 05$^{\mathrm h}$25$\fm$3 & $-$70$^{\circ}$00$\arcmin$ &
    2002-12-27/01:46 & 1.0 & 16 &    \\
LMC Bar 7/P  &                          &                            &
    2002-12-27/02:10 & 0.9 & 17 &    \\
LMC Bar 7/CC &                          &                            &
    2002-12-27/02:34 & 1.0 & 18 &    \\
LMC Bar 7/DD &                          &                            &
    2002-12-27/02:58 & 1.1 & 16 &    \\
LMC Bar 4/G  & 05$^{\mathrm h}$24$\fm$6 & $-$69$^{\circ}$53$\arcmin$ &
    2002-12-27/03:27 & 1.1 & 17 &    \\
NGC 2298     & 06$^{\mathrm h}$49$\fm$0 & $-$36$^{\circ}$00$\arcmin$ &
    2002-12-27/05:15 & 0.9 &  7 & Calibrator: [Fe/H] = $-$1.74 \\
\enddata
\end{deluxetable}
\end{landscape}

\begin{landscape}
\begin{deluxetable}{ccccccccccc}
\tablewidth{0pt}
\tabletypesize{\small}
\tablecaption{Data for Red Giants in the Bar Field \label{tab-dat}}
\tablehead{
\colhead{2MASS ID} &
\colhead{Field} &
\colhead{V} &
\colhead{$\sigma _V$} &
\colhead{I} &
\colhead{$\sigma _I$} &
\colhead{V$_{\sun}$} &
\colhead{$\sigma _{V\sun}$} &
\colhead{$\Sigma$W} &
\colhead{$\sigma _{\Sigma W}$} &
\colhead{Note} \\
\colhead{} &
\colhead{} &
\multicolumn{2}{c}{(mag)} &
\multicolumn{2}{c}{(mag)} &
\multicolumn{2}{c}{(km s$^{-1}$)} &
\multicolumn{2}{c}{(\AA )} &
\colhead{} }
\startdata
05230778-6950057 & 1/A &  17.648 & 0.003 & 16.063
     & 0.007 & 264.8 & 7.6 & 8.55 & 0.12 &  \\
05231484-6950196 & 1/A &  17.486 & 0.004 & 16.120
     & 0.004 & 257.0 & 7.4 & 8.60 & 0.26 &  \\
05225670-6950472 & 1/A &  17.350 & 0.009 & 15.860
     & 0.005 & 282.7 & 7.5 & 8.60 & 0.13 & on N131 HII region \\
05230606-6951113 & 1/A &  17.491 & 0.009 & 15.986
     & 0.005 & 273.9 & 7.6 & 9.01 & 0.12 &  \\
05225436-6951262 & 1/A &  17.349 & 0.008 & 15.954
     & 0.008 & 286.7 & 7.5 & 8.01 & 0.09 & on N131 HII region \\
\enddata
\tablecomments{Table~\ref{tab-dat} is presented in its
entirety in the electronic edition of the Astronomical Journal.
A portion is shown here for guidance regarding its form
and content.}
\end{deluxetable}
\end{landscape}

\begin{deluxetable}{cccccc}
\tablewidth{0pt}
\tablecaption{Derived Quantities for Red Giants in the Bar Field
\label{tab-feh}}
\tablehead{
\colhead{2MASS ID} &
\colhead{[Fe/H]} &
\colhead{$\sigma _{FeH}$} &
\colhead{[$\alpha$/Fe]} &
\colhead{log(Age/yr)} &
\colhead{$\sigma _{logA}$} }
\startdata
05230778-6950057 & -0.29 & 0.14 & -0.25 & 10.00 & 0.17 \\
05231484-6950196 & -0.31 & 0.16 & -0.24 &  9.24 & 0.21 \\
05225670-6950472 & -0.35 & 0.14 & -0.22 &  9.58 & 0.27 \\
05230606-6951113 & -0.16 & 0.14 & -0.32 &  9.43 & 0.25 \\
\enddata
\tablecomments{Table~\ref{tab-feh} is presented in its
entirety in the electronic edition of the Astronomical Journal.
A portion is shown here for guidance regarding its form
and content.}
\end{deluxetable}

\begin{deluxetable}{rlccc}
\tablewidth{0pt}
\tablecaption{Velocity Dispersion vs.\ Metallicity
\label{tab-kin}}
\tablehead{
\multicolumn{2}{c}{Bin} &
\colhead{N$_{\star}$} &
\colhead{$\bar{V}_{\sun}$} &
\colhead{$\sigma_{V_{\sun}}$} \\
\colhead{[Fe/H]$_{\mathrm{min}}$} &
\colhead{[Fe/H]$_{\mathrm{max}}$} &
\colhead{} &
\colhead{(km s$^{-1}$)} &
\colhead{(km s$^{-1}$)}}
\startdata
$-$0.14 & $+$0.14 &  19 & 255.0 & 16.7 $\pm$1.6\\
$-$0.26 & $-$0.14 &  56 & 256.9 & 21.5 $\pm$1.0\\
$-$0.44 & $-$0.26 & 147 & 257.7 & 23.7 $\pm$0.6\\
$-$0.55 & $-$0.44 &  77 & 252.4 & 23.9 $\pm$0.9\\
$-$0.70 & $-$0.55 &  37 & 257.5 & 28.0 $\pm$1.3\\
$-$1.15 & $-$0.70 &  19 & 256.2 & 26.6 $\pm$1.7\\
$-$2.13 & $-$1.15 &  18 & 262.5 & 40.8 $\pm$1.7\\
\enddata
\end{deluxetable}

\begin{deluxetable}{lcccr}
\tablewidth{0pt}
\tablecaption{Metallicities of Stellar Populations in the LMC
\label{tab-pops}}
\tablehead{
\colhead{Population} &
\colhead{Age Estimate} &
\multicolumn{2}{c}{[Fe/H]} &
\colhead{Reference} \\
\colhead{} &
\colhead{(Myr)} &
\colhead{mean} &
\colhead{dispersion} &
\colhead{} }
\startdata
RGB, bar, metal-rich   & 1000--5000    & $-$0.37           & 0.15 & this 
paper
\\
RGB, bar, metal-poor   & $\gtrsim$5000 & $-$1.08           & 0.47 & this 
paper
\\
\hline
B dwarfs               & $<$20         & $-$0.31           & 0.04 &
\citet{rol02} \\
Cepheid variables      & 10--60        & $-$0.34           & 0.15 &
\citet{luc98} \\
Young red giants       & 200--1000     & $-$0.45           & 0.10 &
\citet{smi02} \\
Int.\ age clusters     & 1000--3000    & $-$0.40\tablenotemark{\ddag}
                                                           & 0.22 & OSSH,
\citet{gei97}
                          \\
Planetary nebulae      & 1000--10$^4$  & $-$0.5\tablenotemark{\dag}
                                                           & 0.2  &
\citet{dop97} \\
RR Lyr variables, bar  & $\geq$10$^4$  & $-$1.23\tablenotemark{\ddag}
                                                           & 0.29 &
\citet{gra04} \\
Old clusters           & $\geq$10$^4$  & $-$1.74\tablenotemark{\ddag}
                                                           & 0.36 & OSSH,
\citet{joh04}
                          \\
\enddata
\tablenotetext{\dag}{Average of Ne, Ar, S.}
\tablenotetext{\ddag}{Converted to CG97 scale by Equation~\ref{eqn-cal}.}
\end{deluxetable}

\clearpage

\end{document}